\documentclass[a4paper,11pt]{article}
\pdfoutput=1

\usepackage{jheppub}
\usepackage{customprelude}

\title{Some aspects of symmetry descent}

\author{Iñaki García Etxebarria$^{\clubsuit}$ and}
\author{Saghar S. Hosseini$^{\spadesuit}$}

\affiliation[\clubsuit]{Department of Mathematical Sciences,
  Durham University,\\
  Durham, DH1 3LE, United Kingdom}

\affiliation[\spadesuit]{Kavli Institute for the Physics and
  Mathematics of the Universe (WPI),
  University of Tokyo,\\
  Kashiwa, Chiba 277-8583, Japan}

\emailAdd{inaki.garcia-etxebarria@durham.ac.uk}
\emailAdd{saghar.hosseini@ipmu.jp}

\abstract{In many cases the symmetry structure of quantum field
  theories can be neatly encoded into their associated symmetry
  topological field theory (SymTFT), a topological field theory in one
  dimension higher. For geometrically engineered QFTs in string theory
  this SymTFT has been argued to arise from the background geometry,
  essentially by integration of the topological sector of string
  theory on the horizon of the geometry transverse to the QFT
  locus. In this paper we clarify some subtle aspects of this
  proposal. We take a higher dimensional approach, where the ten
  dimensional string theory fields to be integrated arise as edge
  modes of a topological field theory in eleven dimensions. The
  resulting construction provides a SymTFT generalisation of the
  descent procedure for anomalies.}

\begin{document}
\maketitle
\flushbottom

\section{Introduction}
\label{sec:intro}

A complete definition of any given Quantum Field Theory requires
information both about local dynamics, and about its topological
sector: in general, the same set of local degrees of freedom can be
coupled to different topological sectors. From a modern point of view,
we consider any topological operator in the Quantum Field Theory a
symmetry (in a suitably generalised sense), so an equivalent
restatement of the previous remark is that theories with the same
local dynamics might have different sets of symmetries
\cite{Gaiotto:2010be,Aharony:2013hda,Kapustin:2014gua,Gaiotto:2014kfa}. These
choices of symmetries are often related in simple ways. For instance,
if we have a $d$-dimensional theory $\cT_d$ with a finite symmetry
group $G$,\footnote{Here we can allow for the possibility of
  group-like higher form symmetries, as in \cite{Gaiotto:2014kfa}.} we
can gauge\footnote{Perhaps after tensoring with a topological field
  theory with symmetry $H$. For simplicity of exposition, we will
  also, somewhat imprecisely, refer to this situation as ``gauging''.}
an anomaly-free subgroup $H$ of $G$ and obtain a new theory $\cT_d/H$,
with the same local dynamics but different symmetries. In general $G$
itself need not be anomaly-free, which can lead to some of the
symmetry generators in $\cT_d/H$ to be \emph{non-invertible}, meaning
that for a given operator $\cO$ there is no operator $\cO^{-1}$ in the
theory such that $\cO \cO^{-1}$ is the identity operator.

This is far from an exotic possibility: the existence of such
operators in two dimensions has been understood for many years
\cite{Frohlich:2009gb,Carqueville:2012dk,Brunner:2013xna,Bhardwaj:2017xup,Gaiotto:2019xmp},
and more recently it has been argued, starting with
\cite{Gaiotto:2019xmp,Heidenreich:2021xpr,Choi:2021kmx,Kaidi:2021xfk},
that the same is true in higher dimensions.  We refer the reader to
\cite{Cordova:2022ruw, McGreevy:2022oyu, Freed:2022iao,
  Schafer-Nameki:2023jdn, Brennan:2023mmt, Bhardwaj:2023kri,
  Shao:2023gho} for reviews, and pointers to the extensive
literature. This process can be continued: $\cT_d/H$ will have its own
set of symmetries, and we can gauge a subset of these, leading to a
new theory, and so on. We refer to a choice of representative in this
set of related theories as a choice of global form.

The discussion in the previous paragraph is somewhat unsatisfactory,
in that it started from a theory $\cT_d$ with ``ordinary'' group-like
symmetries, and we accessed more interesting symmetry structures by
sequences of gaugings. But in general, there is no requirement that a
canonical choice for $\cT_d$ exists, and in fact, it is possible that
none of the set of theories related by gauging operations has
group-like symmetries only. A better viewpoint is available
\cite{Ji:2019jhk, Gaiotto:2020iye, Apruzzi:2021nmk, Freed:2022qnc,
  Kaidi:2022cpf, Kaidi:2023maf, Bhardwaj:2023wzd, Baume:2023kkf,
  Brennan:2024fgj, Antinucci:2024zjp, Bonetti:2024cjk,
  Apruzzi:2024htg, DelZotto:2024tae, Bhardwaj:2024qrf,
  Cordova:2024vsq, Nardoni:2024sos, Argurio:2024oym}: instead of
considering each $d$-dimensional theory with the same local dynamics
as $\cT_d$, we study a $(d+1)$-dimensional topological field theory
$\SymTFT_{d+1}$ (which we call the \emph{symmetry topological field
  theory}, or \emph{SymTFT}, as in \cite{Apruzzi:2021nmk}). This
theory admits a gapless boundary condition with the same local
dynamics as $\cT_d$. We denote this gapless boundary condition (which
should be understood as a relative quantum field theory
\cite{Freed:2012bs}) encoding the local dynamics of $\cT_d$ as
$\Lambda(\cT_d)$.  All theories related by gauging of finite
symmetries lead to the same $\Lambda(\cT_d)$. This configuration in
itself should be thought of as a $(d+1)$-dimensional theory on a space
with boundary, but it can be turned into a definition for a
$d$-dimensional field theory with the same local dynamics as $\cT_d$
if $\SymTFT_{d+1}$ admits a gapped interface $\iota_d$ to an anomaly
theory $\cA_{d+1}$.\footnote{That is, an invertible field theory in
  $d+1$ dimensions encoding the anomalies of a (relative) QFT in
  $d$-dimensions.} Pictorially, we have:

\begin{center}
    \includegraphics[width=\textwidth]{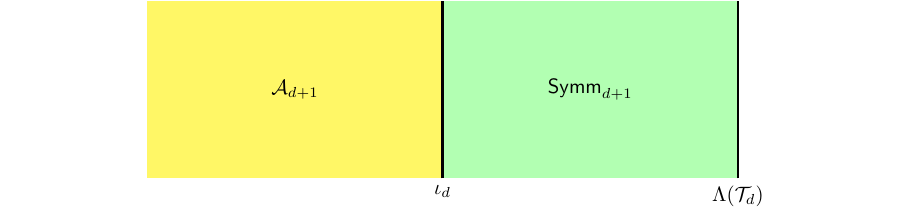}
\end{center}

We obtain a $d$-dimensional theory with anomaly $\cA_{d+1}$ by
colliding $\iota_d$ with $\Lambda(\cT_d)$. Different choices of global
form for $\cT_d$ correspond to different choices of the pair
$(\cA_{d+1},\iota_d)$: some of the topological operators in
$\SymTFT_{d+1}$ will become trivial when we collapse $\iota_d$ and
$\Lambda(\cT_d)$, but some will survive as symmetries of the resulting
field theory. In this way, we have reformulated the problem of
understanding all theories related to $\cT_d$ and their symmetries
into two parts: first determining $\SymTFT_{d+1}$ given $\cT$ (or
$\Lambda(\cT_d)$) and then classifying the pairs $(\cA_{d+1},\iota_d)$
that $\SymTFT_{d+1}$ can attach to.

\medskip

Our focus in this paper will be on supersymmetric quantum field
theories obtained by placing string theory or M-theory at isolated
conical singularities of the form $C(L)$, with $L$ the base of the
cone. This class of configurations leads to superconformal field
theories living at the singular base of the cone. Generically these
superconformal theories do not admit any weakly coupled description,
so we have very poor knowledge of $\Lambda(\cT_d)$. Nevertheless, it
was argued in \cite{Apruzzi:2021nmk} (see also
\cite{Bah:2019rgq,Bah:2020jas} for related earlier works) that
$\SymTFT_{d+1}$ can be obtained (in the M-theory case) by performing a
reduction of the topological terms in the M-theory action on $L$,
which reduces the computation to a somewhat technical but fully
solvable problem in algebraic topology.\footnote{Ideally we would like
  to extract $\SymTFT_{d+1}$ from the geometry as a fully extended
  topological quantum field theory, but at the moment it is only known
  how to systematically extract the information described in the
  text.}

Our main goal in this paper is to clarify one issue in the analysis of
\cite{Apruzzi:2021nmk} that remained somewhat puzzling: while the
terms in $\SymTFT_{d+1}$ related to anomalies were computed via a
straightforward integration on $L$, there was also a $BF$ sector,
schematically of the form
\[
  S_{BF}(B,A)=  2\pi i N \int B \wedge dA  \, ,
\]
which was computed using entirely different (and somewhat indirect)
methods. Our goal in this note is to bring the two viewpoints closer
together by explaining how this $BF$ theory can be derived by
integration on $L$ of an auxiliary theory in one dimension higher.

\medskip

Let us briefly review how \cite{Apruzzi:2021nmk} argues for the
existence of a $BF$ sector in $\SymTFT_{d+1}$. The basic setup is
string theory or M-theory\footnote{For reasons that will become
  apparent, we are currently not able to satisfactorily apply our
  techniques to M-theory. We are hopeful that the difficulties will be
  surmountable, but we do not know how to do so at the moment.} on a
Calabi-Yau cone $C(L)$ with Sasaki-Einstein base $L$ times a manifold
$\cM^{d}$:
\begin{center}
    \includegraphics[width=\textwidth]{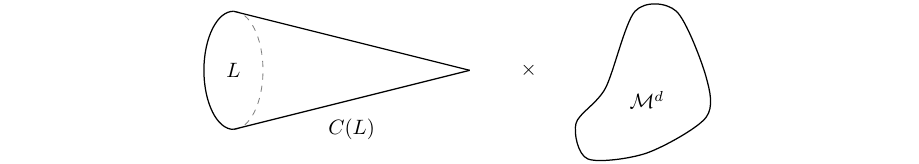}
\end{center}
In M-theory $\dim(L)=10-d$ and in type II string theory
$\dim(L)=9-d$. We will assume that the singular locus of $C(L)$
consists of an isolated singularity at the tip of the cone, and that
$\cM^d$ is spin and torsion-free. There are light degrees of freedom
living at the singular locus of $C(L)$, which in the examples studied
in \cite{Apruzzi:2021nmk} lead to a $d$-dimensional SCFT living on
$\cM^d$. As argued in \cite{GarciaEtxebarria:2019caf}, the information
given so far defines only the local data of the SCFT (namely, the
relative theory we have denoted by $\Lambda(\cT_d)$ in the
introduction) and the rest of the data necessary for fully defining a
SCFT $\cT_d$ are instead encoded in a choice of boundary condition at
infinity for the supergravity fields on $\cM^d\times C(L)$.

One of the main results of \cite{Apruzzi:2021nmk} is to reconcile this
picture, which is very natural from a string theory point of view,
with the SymTFT viewpoint described in the introduction. The basic
idea is that $\SymTFT_{d+1}$ arises from reducing the topological
sector of string theory over the base of the cone $C(L)$:
\begin{center}
    \includegraphics[width=\textwidth]{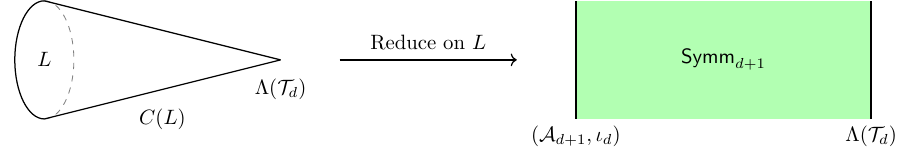}
\end{center}

In this proposal, we identify the tip of the cone with the gapless
boundary $\Lambda(\cT_d)$, and the boundary conditions at infinity
with the pair $(\cA_{d+1},\iota_d)$.\footnote{\label{ft:strings-QFT
    map}Although the details of correspondence need to be worked out,
  the implicit expectation in \cite{GarciaEtxebarria:2019caf} is that
  the identification should go as follows. A choice of boundary
  conditions, given by the expectation value of a maximal isotropic
  subgroup of fluxes at infinity, as described in
  \cite{GarciaEtxebarria:2019caf}, corresponds to a choice of
  $\iota_d$. The partition function of the resulting string theory
  configuration is not well defined as a number, but it is rather a
  section of a line bundle over the space of boundary conditions over
  the boundary $\cM^d\times L$. Restrict the supergravity fields to be
  flat asymptotically. A subset of such fields will be given by flat
  representatives of torsional cohomology groups on $L$ times
  representatives of integral cohomology classes on $\cM^d$. Using the
  Künneth formula, which is an isomorphism in this case, these
  supergravity backgrounds can be canonically identified with
  backgrounds for the symmetries of $\cT_d$ on $\cM^d$. (We assume
  that there is no torsion in the cohomology of $\cM^d$ here, see
  \cite{GarciaEtxebarria:2019caf} for details.) The change in the
  partition function of string theory under gauge transformations
  within this subset of fields is precisely given by the anomaly
  theory $\cA_{d+1}$.}

It was shown in \cite{Apruzzi:2021nmk} that a subsector of
$\SymTFT_{d+1}$ (leading to $\cA_{d+1}$ for suitable choices of
boundary conditions) arises quite naturally from integrating the
Chern-Simons term in the M-theory action over $L$.

The $BF$ sector in $\SymTFT_{d+1}$ is more subtle. For concreteness we
consider, as in \cite{Apruzzi:2021nmk}, the case of $d=5$ SCFTs
arising from putting M-theory on singular Calabi-Yau threefolds. As
studied in \cite{Morrison:2020ool,Albertini:2020mdx}, the local
dynamics for these SCFTs can be completed to theories with either
1-form or 2-form symmetries, and the boundary conditions for the $BF$
subsector of $\SymTFT_{6}$ determine which kind of symmetries we
have. The 1-form symmetries have generators coming from $G_4$ fluxes
and the 2-form symmetries from the $G_7$ fluxes (under the kind of
identification reviewed in footnote~\ref{ft:strings-QFT map}, and
explained in more detail in \cite{GarciaEtxebarria:2019caf}.  These
flux generators do not commute with one another \cite{Moore:2004jv,
  Freed:2006ya, Freed:2006yc}, and as a result, when choosing boundary
conditions we need to choose to which of $G_4$ or $G_7$ we give
Dirichlet boundary conditions, we cannot give Dirichlet boundary
conditions to both. In terms of the field theory we have that we
cannot realise both the 1-form and 2-form symmetry at the same
time. This hints at the fact that if the $BF$ action was to be
obtained from a reduction of a supergravity action, it should be one
that contains both the $G_4$ and $G_7$ fluxes. Ignoring some (crucial)
complications from the presence of the Chern-Simons term, the closest
object is the kinetic term $\int G_4 \wedge *G_4$, if one observes
that $G_4$ and $G_7$ fluxes are dual to one another in M-theory. This
expectation is borne out by the analysis in \cite{Apruzzi:2021nmk},
which constructs the algebra of topological operators in the $BF$
theory from the reduction on $L$ of the algebra of topological
operators in the M-theory background.

The discussion above explains why it is subtle to obtain the expected
$BF$ theory from a dimensional reduction: to reproduce the
non-commutativity we would need an action that simultaneously
describes electric ($G_4=dC_3$) and magnetic ($G_7=dC_6$) fluxes. This
is closely related, by viewing the pair $(C_3, C_6)$ as a
two-component field, to the famously difficult problem of writing an
action for a self-dual field.

\begin{figure}
  \centering
  \includegraphics{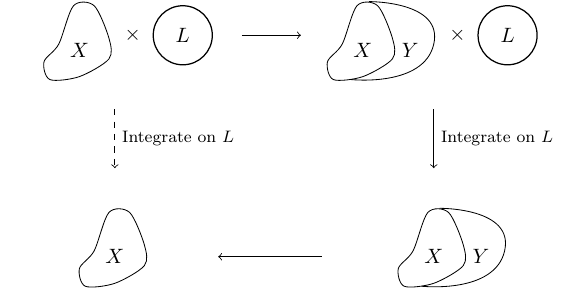}
  \caption{We want to understand the emergence of the $BF$ theory on
    $X$ from compactification of self-dual dynamics on $L$. (Dashed
    arrow on the left hand side of the diagram.) We do this by
    promoting the theory on $X\times L$ to a Chern-Simons theory on
    $Y\times L$, where $\partial Y=X$, integrating this Chern-Simons
    theory on $L$, and restricting back to $X$.}
  \label{fig:approach-summary}
\end{figure}

There are multiple ways of approaching the problem of constructing
actions, or directly the partition function, for a self-dual field
\cite{Witten:1996hc, Witten:1998wy,Witten:1999vg,Moore:1999gb,Freed:2000ta,Gukov:2004id,Belov:2004ht,Belov:2006jd, Belov:2006xj}. In this paper, we approach this problem using a
strategy initiated by Witten \cite{Witten:1996hc}, in which we define
the partition function of a self-dual field on $X$ in terms of
Chern-Simon theory on $Y$, with $\partial Y = X$. Suitable choices of
boundary conditions for the Chern-Simons field (which we discuss in
detail below, following \cite{Belov:2004ht,Hsieh:2020jpj}) lead to the
emergence of the self-dual degrees of freedom on $X$. From this point
of view, our task decomposes into two (simpler) tasks: first we need
to reduce the Chern-Simons action on our internal manifold $L$, and
then we need to understand how the effective theory arising from
reduction on $L$ behaves on a manifold with boundary. This is
summarised in figure~\ref{fig:approach-summary}.

The central result of our analysis is stated in a precise way in
eq.~\eqref{eq:symm-inflow} below. Intuitively, what that equation is
saying is that there is a form of inflow for the SymTFT:
\[
  \label{eq:symm-inflow-simplified}
  \Delta \int_L \cL_{BF} = \delta \cL_{\SymTFT}
\]
where $\cL_{BF}$ and $\cL_{\SymTFT}$ are suitable notions of
``Lagrangian densities'' for the $BF$ and SymTFT, $\Delta$ denotes
gauge transformations, and $\delta$ denotes the exterior
derivative. The formalism of Hopkins and Singer \cite{Hopkins:2002rd}
is very useful in making the notion of gauge transformations of
Lagrangian densities precise in topologically non-trivial
configurations, we review the basics in
section~\ref{sec:Hopkins-Singer}. There are two key advantages of using this formalism. The
first one is that it allows us to treat in a systematic way
topologically non-trivial bundles. In particular, we will be
interested in topologically non-trivial flat bundles. While it is
possible to treat such bundles using the language of differential
forms, for example following the ideas in \cite{Camara:2011jg}, we
find the language of differential cochains both better founded
mathematically, and easier to work with in practice.\footnote{A
  question where the differential cohomology formalism is useful is
  the fundamental problem of computing linking pairings between
  torsion cycles, which to our knowledge does not have an easily
  computable answer using the differential form formalism of
  \cite{Camara:2011jg}. See \cite{Casas:2023wlo} for recent progress
  in this direction.} The second key advantage of this formalism,
compared with the language of differential characters used in
\cite{Apruzzi:2021nmk}, for instance, is that it allows us to keep
control of the topological aspects of the problem in a way that is
local. (In physics terms, we will be producing Lagrangian densities,
and not just actions.) This fact allows us to make our formalism
actually useful in practice: otherwise, since we are viewing the
SymTFT as arising from boundary modes, it would be difficult to place
it on spaces with boundary!  

We expect that a modification of our approach (replacing ordinary
differential cohomology with twisted differential $K$-theory) will
allow us to recover not only the $BF$ sector but also the anomaly
theory, at least in cases where the mathematical formalism is
understood well enough. We provide evidence for this expectation in
section~\ref{sec:couplings}.

\medskip

\emph{A note on related recent literature:} The papers
\cite{vanBeest:2022fss,Apruzzi:2022rei,Lawrie:2023tdz,Bah:2023ymy,Apruzzi:2023uma,Baume:2023kkf,Yu:2023nyn,
  Basile:2023zng} appeared during the (fairly long) preparation of the
results we present here.\footnote{See also \cite{Casas:2023wlo} for a
  recent analysis of the emergence of the $BF$ theories from a
  different perspective.} They include a discussion of ideas related
to those in this paper for the derivation of the $BF$ action, among
many other interesting results, and in particular they also start from
a theory in eleven dimensions (in the string theory case) and propose
that the SymTFT can be understood from dimensional reduction of the
eleven dimensional theory. Nevertheless, we believe that the analysis
in this paper is still useful, as it clarifies many of the subtleties,
both technical and conceptual, which we encountered in making this
picture precise and which were not addressed in the works just
mentioned. Our new results,
compared to the results given in those papers, are that we will
explain in detail which boundary conditions one needs to take in the
eleven dimensional theory, how the edge modes --- or in other words,
the fields in the SymTFT --- relate to the dynamical fields in eleven
dimensional theory, and we will explain various subtle aspects of the
mathematics involved in the case of topologically non-trivial
geometries. With these results in hand we will be able to give a
precise meaning to, and prove, our main new
result~\eqref{eq:symm-inflow-simplified}.

\section{Differential cohomology}
\label{sec:Hopkins-Singer}

Our starting point is abelian Chern-Simons theory at level $k$ in
$4n+3$-dimensions. The action $S$ for this theory can be written as
$S=2\pi i\, \CS_k[A]$, with (heuristically)
\[
  \label{eq:CS-heuristic}
  \CS_k[A] = \frac{k}{2} \int A\wedge dA\, ,
\]
where $A$ is a $2n+1$ form. The most familiar case is $n=0$, where the
self-dual field on the boundary is a self-dual
boson. Eq.~\eqref{eq:CS-heuristic} is heuristic for three reasons.
First, we are using differential form notation for the field $A$, but
in the cases of interest to us $A$ cannot be globally defined as a
differential form, but it is rather a connection on a topologically
non-trivial bundle. The second aspect of~\eqref{eq:CS-heuristic} that
needs clarification is that, in fact, we will be interested mostly in
the case of $A$ being a \emph{flat} connection on a topologically
non-trivial bundle. (Our discussion is about IR effects, and non-flat
modes in the same topological sector encode massive excitations, which
we want to integrate out.) So, naively, $dA=0$. These two
complications can be dealt with by switching to the language of
differential cohomology. Below we will give a very brief review of the
main aspects of this formalism as they apply to our case. A last
subtlety in the interpretation of~\eqref{eq:CS-heuristic} concerns the
quantisation of $k$. In general, this Chern-Simons theory makes sense
for arbitrary $k\in\bZ$, but for odd values of $k$ there is an
implicit dependence on the Wu structure of the manifold. In the
(oriented) $n=0$ case the Wu structure reduces to the spin structure,
and we have the familiar statement that Chern-Simons theory at odd
values of $k$ (or half-integral values, depending on conventions)
depends on the spin structure. We will address this final subtlety
below, after introducing some formalism we need for addressing the
first two points.

Since we will need to consider a refinement of the connection $A$
itself, and not just the Chern-Simons action, we will adopt the
formalism of Hopkins–Singer \cite{Hopkins:2002rd}.\footnote{We will be
  working mostly with ordinary cohomology, where an equivalent cochain
  formalism was already introduced by Cheegers and Simons in
  \cite{10.1007/BFb0075216}. The formalism of Hopkins and Singer
  allows for studying differential cochains in generalised cohomology
  theories, so with an eye towards generalisations we will refer to it
  as the Hopkins-Singer formalism.}  Our goal here is simply to set
notation and review the basic operational rules. We encourage the
interested reader to read
\cite{Freed:2000ta,Hopkins:2002rd,Hsieh:2020jpj} for in-depth
discussions.

\medskip

Consider the cochain complex $\{\diff C(l)^\bullet(\cM),d\}$ with
\begin{equation}
  \diff C(l)^p(\cM)=
        \begin{cases}
            C^p(\cM;\bZ)\times C^{p-1}(\cM;\bR)\times \Omega^p(\cM;\bR) & \text{for}\quad p\geq l \\
            C^p(\cM;\bZ)\times C^{p-1}(\cM;\bR) & \text{for}\quad p<l
        \end{cases}
\end{equation}
and differential
\begin{equation} \label{eq:differential}
    \begin{split}
        d(c,h,\omega)=&\, (\delta c,\omega-c-\delta h,d\omega)\quad \text{for}\quad  (c,h,\omega)\in  \diff C(l)^{p}(\cM)\\
        d(c,h)=&
            \begin{cases}
                (\delta c,-c-\delta h,0) &\text{for}\quad (c,h)\in \diff C(p)^{p-1}(\cM)\\
                (\delta c,-c-\delta h) &\text{otherwise.}
            \end{cases}
    \end{split}
\end{equation}
We can alternatively think of elements of $\diff C(l)^p(\cM)$ for
$p<l$ as triples $(c, h, \omega)$ with $\omega=0$. We call elements of
$\diff C(l)^p(\cM)$ \emph{differential cochains}, and define
\emph{differential cocycles} as the closed differential cochains:
\begin{equation}\label{eq:diffco}
    \diff Z(l)^p(\cM) \df \{\diff x\in \diff C(l)^p(\cM) \mid d\diff x=0\}\, .
\end{equation}

For notational convenience, we introduce maps $I$, $\sh$ and $R$ such
that for the differential cochain
$\diff a=(c,h,\omega)\in \diff C(l)^p(\cM)$
\[\label{eq: components}
  \big(I(\diff a),\sh(\diff a),R(\diff a)\big)=(c,h,\omega)\, ,
\]
and define\footnote{This definition is a minor deviation from the one
  in \cite{Hopkins:2002rd}, but we find it slightly more convenient.}
$\diff C^p(\cM)\df \diff C(p)^p(\cM)$,
$\diff Z^p(\cM)\df \diff Z(p)^p(\cM)$. The names of the maps refer to the fact that for a
differential cocycle $\diff a$, $c=I(\diff a)$ gives a cocycle
representing the characteristic class of the associated bundle,
$h=\sh(\diff a)$ represents an extension of its holonomy to $\bR$
(see~\eqref{eq:Cheeger-Simons-chi} below), and $\omega=R(\diff a)$ is
its curvature.

Finally, we note for future reference that the condition $d\diff a=\diff 0$
for a cocycle $\diff a\in \diff Z(l)^p(\cM)$ implies that its
components satisfy $\delta \sh(\diff a) = R(\diff a) - I(\diff a)$.

The differential cohomology group $\diff H(l)^p(\cM)$ is then obtained
in the familiar way:
\begin{equation}
  \diff H(l)^p(\cM) \df \diff Z(l)^p(\cM)/ d \diff C(l)^{p-1}(\cM)\, .
\end{equation}
The special case $\diff H(p)^p(\cM)$ coincides with the Cheeger-Simons
differential cohomology group $\diff H^p(\cM)$
\cite{10.1007/BFb0075216}. In particular, the Cheeger-Simons
differential character $\chi([\diff x])\in \bR/\bZ$ is given by
\[
  \label{eq:Cheeger-Simons-chi}
  \chi([\diff x]) = \sh(\diff x) \mod 1\, .
\]
Here (and below) we denote elements of $\diff H^p(\cM)$ by
$[\diff x]$, where $\diff x\in \diff Z(p)^p(\cM)$ is some
representative of the differential cohomology class, defined up to an
exact cocycle.

\medskip

The next ingredient we need is a notion of a product between
differential cochains, generalising the cup product in differential
cohomology. Given two differential cochains
$\diff a_1=(c_1,h_1,\omega_1)$ and $\diff a_2=(c_2,h_2,\omega_2)$
their product $\diff a_1 \cdot \diff a_2$ is a new cochain with
components
\begin{equation} \label{eq:product} (c_1 \cup c_2,(-1)^{|c_1|} c_1
  \cup h_2 + h_1 \cup \omega_2 + B(\omega_1, \omega_2), \omega_1
  \wedge \omega_2) \, .
\end{equation}
For a cochain $x$, we denote its degree as $|x|$. Here  $B$ is any natural
homotopy between $\wedge$ and $\cup$
\begin{equation}\label{homotopy}
  \delta(B(\omega_1,
  \omega_2))+B(d\omega_1, \omega_2)+(-1)^{|\omega_1|}B(\omega_1,
  d\omega_2) =\omega_1\wedge \omega_2-\omega_1\cup \omega_2\, .
\end{equation}
where we are promoting differential forms to cochains as needed. Note
that whenever $\omega_1=0$ or $\omega_2=0$ we can choose
$B(\omega_1,\omega_2)=0$ (since the right-hand side
of~\eqref{homotopy} vanishes and this choice of homotopy is certainly
natural; or alternatively using the explicit expression given in
\cite{10.1007/BFb0075216}). A straightforward computation shows that
for $\diff a \in \diff C(p)^q(\cM)$, $\diff b \in \diff C(r)^s(\cM)$
we have
\[
  \label{eq:adb}
  d(\diff a \cdot \diff b) = (d\diff a)\cdot\diff b + (-1)^q \diff a\cdot d(\diff b)\, ,
\]
In particular this implies that the product defined above induces a
product in differential cohomology classes, which is precisely the
product defined by Cheeger and Simons \cite{10.1007/BFb0075216}. We
note that, for $\diff a$ and $\diff b$ cocycles, we have that
$\diff a\cdot \diff b$ and
$(-1)^{\deg(\diff a)\deg(\diff b)}\diff b\cdot \diff a$ are equivalent
up to gauge transformations, where $\deg(\diff x) \df p$
for any $\diff x\in\diff C(l)^p(\cM)$.\footnote{This may be shown
  using the fact that cup product on cocyles is graded commutative up
  to the coboundary of cup-1 product together with the definition of
  $B(\omega_1,\omega_2)$ given in \cite{10.1007/BFb0075216} in terms
  of a sum of the cup product evaluated on subdivisions:
  $B(\omega_1, \omega_2) - B(\omega_2, \omega_1) = -\sum_i (\omega_1
  \cup \omega_2 - \omega_2 \cup \omega_1)(...)$.}

\medskip

Given a fibre bundle with closed oriented fibres it is possible to
define a notion of integration along the fibre for differential
cochains \cite{Hopkins:2002rd}. In this note, we are only interested
in the case of trivial fibrations, namely spaces of the form
$\cM=B\times \Phi$, where the fibre\footnote{We denote the fibre as
  $\Phi$ rather than $F$, to avoid confusion with the field strength
  $F$.  } $\Phi$ is $n$-dimensional, closed and oriented.\footnote{For
  cohomology theories $\diff H$ we need $\Phi$ to be
  $\diff H$-oriented \cite{Hopkins:2002rd}. In particular, if
  $\diff H$ is differential complex $K$-theory, we want $\Phi$ to
  admit a spin$^c$ structure, which is always the case for any
  oriented three-manifold, the basic class of examples considered in
  this paper.} In this case, we define the integration map
\[
  \int_\Phi \colon \diff C(p)^p(B\times \Phi) \to \diff C(p-n)^{p - n}(B)
\]
by
\begin{equation}\label{eq:IntDef}
  \int_\Phi (c,h,\omega) \df \left(c/\Phi, h/\Phi, \omega/\Phi \right)
\end{equation}
where on the slant products \cite{spanier1989algebraic} on the right
hand side we have abused notation (as we will keep doing in this
paper), and denoted by $\Phi\in Z_n(\Phi)$ the fundamental class of
$\Phi$. Given a cochain $c\in C^l(B\times \Phi; A)$, for $A$ an abelian
group, and a chain $v\in C_p(\Phi;A)$ the (bilinear) slant product
$c/v \in C^{l-p}(B;A)$ satisfies $(c/v)(u)=c(u\times v)$ for all
$u\in C_{l-p}(B;A)$. A property of the slant product that we will need
later is
\[
  \label{eq:slant-differential}
  \delta(c/v) = (\delta c)/v + (-1)^{l-p} c / (\delta v)\, ,
\]
so the slant product on (co)chains descends to (co)homology.
Additionally, viewing $\omega$ as a differential form, $\omega/\Phi$
coincides with the usual notion of integration along $\Phi$. Motivated by
this, we will sometimes abuse notation and write $x/\Phi$ as $\int_\Phi x$,
even when $x$ is a cochain.

\medskip

So far we have assumed that the fibre is closed. In case the fibre has
a boundary the discussion above still goes through, with some
modifications described in detail in \cite{Hopkins:2002rd}. A
particularly important result, in this case, is the following version
of Stokes' theorem for $\diff x \in \diff C(p)^q(\cM)$:
\begin{equation}
  \label{eq:Stokes}
  d\int_\Phi \diff x = \int_\Phi d \diff x + (-1)^{|\diff x|-\dim(\Phi)} \int_{\partial \Phi} \diff x\, ,
\end{equation}
which follows from a short computation
using~\eqref{eq:slant-differential}.  If $\Phi$ is closed then this
version of Stokes' theorem implies that integration descends to
differential cohomology
\[
  \int_\Phi \colon \diff H^p(B\times \Phi) \to \diff H^{p-\dim(\Phi)}(B)
\]
in the obvious way.

\medskip

Before we proceed any further, let us briefly discuss a few simple
examples that illuminate these techniques, and which play an important
role below. In all cases we take the base $B$ to be a point, which we
denote by ``$\pt$'', so our total manifold is
$\cM\df \pt\times \Phi\cong \Phi$.

\subsubsection*{Example 1: $U(1)$ theory in two dimensions}

Consider first the case of an ordinary $U(1)$ gauge 1-form connection
on a two dimensional manifold $M$. We take the action to be
$2\pi i\int_M F$ (up to a small refinement we clarify
momentarily). This theory is trivial whenever $M$ is closed, but it
can be useful when studying Wilson lines on $\partial M$, for
instance.
We will find it useful to reformulate the action of this theory in the
language of differential cohomology as
\[
  \label{eq:diff-cohom-action}
  S_{2d}[\diff a] = 2\pi i\, \sh\left(d\int_{M} \diff a\right)\, ,
\]
with $\diff a\in \diff Z^2(M)$. (By the notation ``$S_{2d}[\diff a]$'' we
mean an action $S_{2d}$ depending functionally on the cocycle $\diff a$.)
Equivalently, since the action appears in the form
$e^{-S_{2d}[\diff a]}$ in the path integral, we can write,
using~\eqref{eq:Cheeger-Simons-chi}:
\[
  \label{eq:F-holonomy}
  \frac{1}{2\pi i}S_{2d}[\diff a] = \chi\left(\left[d\int_M \diff a\right]\right) \mod 1\, .
\]
Note that generically $R(\int_M\diff a)=\int_M R(\diff a)\neq 0$, so
despite the notation $d\int_M\diff a$ is not pure gauge, or
equivalently $\left[d\int_M\diff a\right]\in \diff H(1)^1(\pt)$ does
not necessarily vanish, and therefore the
holonomy~\eqref{eq:F-holonomy} is not necessarily trivial.

\medskip

To see that this gives the action we are after, compute
\[
  \label{eq:M-integral}
  d\int_{M} \diff a  = d\, (I(\diff a)/M, 0, R(\diff a)/M) = (0, (R(\diff a) - I(\diff a))/M, 0)
\]
where on the first equality we have used that $h(\diff a)/M=0$ for
degree reasons, and similarly
$\delta I(\diff a)/M = \delta R(\diff a)/M=0$ on the second one. Since
$I(\diff a)$ is an integral cochain we find
\[
  \frac{1}{2\pi i}S_{2d}[\diff a] = \sh\left(d\int_{M} \diff a\right) = (R(\diff a) - I(\diff a))/M = \int_M F \mod 1
\]
as desired, where $F\df R(\diff a)$.

To see why~\eqref{eq:diff-cohom-action} is useful, take $M$ to have
non-vanishing boundary $\partial M$. By applying~\eqref{eq:Stokes} we
immediately get
\[
  \label{eq:2d-boundary-holonomy}
  S_{2d}[\diff a] = 2\pi i\, \sh\left(\int_{\partial M} \diff a\right)\, .
\]
So we end up with the holonomy on the boundary, as expected. Note that
we have not yet quotiented by gauge transformations, so this holonomy
lives in $\bR$. We can also obtain the same result starting
from~\eqref{eq:M-integral} and using the fact that $\diff a$ is a
cocycle (so $d\diff a=\diff 0$), which then implies
$R(\diff a) - I(\diff a) = \delta \sh(\diff a)$,
so~\eqref{eq:M-integral} gives
$S_{2d}[\diff a] = 2\pi i\, \delta \sh(\diff
a)/M$. Using~\eqref{eq:slant-differential} and the fact that
$\sh(\diff a)/M=0$ for degree reasons, we get
$S_{2d}[\diff a] = 2\pi i\, \sh(\diff a)/\partial M$, which agrees
with~\eqref{eq:2d-boundary-holonomy}.

\medskip

Consider now gauge transformations, which act on $\diff a$ as
$\diff a\to \diff a + d\diff \lambda$, with
$\diff \lambda\in \diff C(2)^1(M)$. Parameterising
$\diff \lambda=(c, f, 0)$, we have
$d\diff \lambda=(\delta c, - c - \delta f, 0)$, with $f\in C^0(M;\bR)$
and $c\in C^1(M;\bZ)$. This gauge transformation acts
on~\eqref{eq:M-integral} by
\[
  d\int_M \diff a \to d\int_M \diff a + d\int_M d\diff \lambda =
  d\int_M\diff a - \int_{\partial M} d\diff\lambda & = (0,(\sh(\diff
  a)+\delta f + c)/\partial M, 0)\\
  & = (0,(\sh(\diff a) + c)/\partial M, 0) \, .
\]
We recognise the term $\delta f$ as the one generating small gauge
transformations. These do not affect the value of this integral (since
$\delta f/\partial M=\delta(f/\partial M)=0$ for degree reasons). On
the other hand, $c$ is the part that generates large gauge
transformations on the boundary, and such transformations do change
the value of the action. Since $c$ is an integral cochain, the change
is by integer multiples of $2\pi i$, so it does not affect the
physics.

\subsubsection*{Example 2: $U(1)\times U(1)$ theory in even dimensions}

As a slightly more elaborate version of this example, consider the
case where $B$ is still a point, which we denote by ``$\pt$'',
$\dim(\Phi)=2n$, and we want to define a theory with action
\[
  \label{eq:12-action}
  S_k \df 2\pi i\, k \int_\Phi F_1\wedge F_2\, ,
\]
where $F_1$ and $F_2$ are field strengths of degree $n$ for abelian
higher form fields. An important application of such an action is in
defining a discrete gauge $\bZ_k$ theory \cite{Banks:2010zn} on
$\partial \Phi$ in terms of a bulk theory on $\Phi$. (See also example 3
below.) As in the previous example, we reformulate this action in
differential cohomology as
\[
\label{eq:U(1)U(1)}
  S_k = 2\pi i\, k\, \sh\left(d\int_\Phi \diff a_1\cdot \diff a_2\right)\, ,
\]
where $\diff a_1,\diff a_2\in \diff Z^n(\Phi)$. The integral
of $\sh(\diff a_1 \cdot \diff a_2)\in C^{2n-1}(\Phi;\bR)$ over $\Phi$
vanishes for dimensional reasons and therefore
\begin{equation}\label{eq:aa}
  \int_\Phi \diff a_1 \cdot \diff a_2 = (I(\diff a_1 \cdot \diff a_2)/\Phi, 0, R(\diff a_2 \cdot \diff a_2)/\Phi)
  \in \diff C(0)^0(\pt) \cong \bZ \times\bR\, .
\end{equation}

\medskip

Assume first that $\partial \Phi=0$. Then by Stokes'
formula~\eqref{eq:Stokes} the result of integration is actually closed
under $d$, which implies the familiar relation
$I(\diff a_1\cdot \diff a_2)/\Phi = R(\diff a_1 \cdot \diff a_2)/\Phi$.

\medskip

More generally, if $\Phi$ has a boundary,~\eqref{eq:Stokes} gives
(similarly to the previous example)
\[
  \label{eq:12-boundary}
  d\int_\Phi \diff a_1 \cdot \diff a_2 & = d(I(\diff a_1 \cdot \diff a_2)/\Phi, 0, R(\diff a_2 \cdot \diff a_2)/\Phi) \\
  & = (0, (R(\diff a_1\cdot \diff a_2) - I(\diff a_1\cdot \diff a_2))/\Phi, 0)\\
  & = (0, \sh(\diff a_1\cdot \diff a_2)/\partial \Phi, 0)\, .
\]
Coming back to~\eqref{eq:12-boundary}, we can rewrite it in more
familiar notation:
\[
  \label{eq:integral-Stokes}
  \int_{\Phi} R(\diff a_1)\wedge R(\diff a_2) - \int_\Phi I(\diff a_1) \cup I(\diff a_2) = \int_{\partial \Phi} \sh(\diff a_1\cdot \diff a_2)\, .
\]
Assuming that $I(\diff a_1)$ is topologically trivial, then there is a
globally defined connection $A_1\in \Omega^1(\Phi)$, and the equation can be written as
\[
  \int_\Phi F_1\wedge F_2 - \int_{\Phi} I(\diff a_1) \cup I(\diff a_2) = \int_{\partial \Phi} A_1\wedge F_2
\]
where we have denoted the field strength $R(\diff a_i)$ by
$F_i$. Gauge transformations shift the right hand side and left hand
sides by (equal) integers, so this equation is often written as
\[
  \int_\Phi F_1\wedge F_2 = \int_{\partial \Phi} A_1\wedge F_2 \mod 1\, .
\]
In our derivation below the more abstract
version~\eqref{eq:integral-Stokes} of this equation will play a key
role.

\medskip

It is an illuminating (and important for our later purposes) exercise
to check the variation of the integral~\eqref{eq:aa} under gauge
transformations of $\diff a_i$. The latter are given by the ambiguity
of $\diff a_i$ by an exact cocycle $d\diff b_i\in \diff Z^{n}(\Phi)$
with $\diff b_i\in\diff C(n)^{n-1}(\Phi)$.  Under these gauge
transformations $\diff a_i \to \diff a_i+d\diff b_i$ and
(using~\eqref{eq:adb})
\[
  \label{eq:transaa}
  \diff a_1\cdot \diff a_2 \to 
  (\diff a_1+d\diff b_1)\cdot(\diff a_2+d\diff b_2)
    = \diff a_1\cdot \diff a_2+d \diff \lambda
    \, ,
\]
with
$\diff \lambda\df (-1)^n \diff a_1\cdot \diff b_2+\diff b_1\cdot \diff
a_2+(-1)^n\diff d b_1\cdot \diff b_2$. (There is some ambiguity in the
choice of $\diff\lambda$, here we pick a representative that
simplifies some of the formulas later on.) Note that
$R(\diff \lambda)=0$, since $R(\diff b_i)=0$.  Thus, if
$\partial \Phi\neq 0$, then by~\eqref{eq:Stokes} the gauge transformation
results in the boundary term
\begin{equation}
  \int_{\partial \Phi}\diff \lambda=(I(\diff \lambda)/\partial \Phi, 0, R(\diff \lambda)/\partial \Phi)
\end{equation}
where again the integral $\sh(\diff \lambda)/\partial \Phi$ for
$\sh(\diff \lambda)\in C^{2n-2}(\Phi;\bR)$ vanishes for dimensional
reasons. In components, let $\diff a_i=(c_i,h_i,w_i)$,
$\diff b_i=(n_i,r_i,0)$ and so
$d\diff b_i=(\delta n_i,-n_i-\delta r_i,0)$. Then,
\[
  \label{eq:boundarymode-BF}
  I(\diff \lambda)/\partial \Phi& =I((-1)^n\diff a_1\cdot \diff b_2+\diff b_1\cdot \diff a_2+(-1)^nd\diff b_1\cdot \diff b_2)/\partial \Phi\\
  & =\int_{\partial \Phi}
    (-1)^n c_1\cup n_2+n_1\cup c_2+(-1)^n \delta n_1\cup  n_2\, ,
\]
and $R(\diff \lambda)/ \partial \Phi=0$. From here
\[
  S_k \to S_k + 2\pi i\, k\, \sh\left(d\int_{\Phi}d\diff \lambda\right) & = S_k + 2\pi i\, k\,\sh\left(d\int_{\partial \Phi} \diff \lambda\right) \\ &= S_k - 2\pi i\, k \int_{\partial \Phi}
  (-1)^n c_1\cup n_2+n_1\cup c_2+(-1)^n \delta n_1\cup n_2\, .
\]
Assuming $k\in\bZ$ the variation is therefore an integer, and
$e^{-S_k}$ is gauge invariant.

\subsubsection*{Example 3: $BF$ theory on a space with boundary}

As our final example, we consider the theory with action
(schematically)
\[
  S_{k,BF} = 2\pi i\, k\int_\Phi A_1\wedge F_2\, .
\]
This action (which models a $\bZ_k$ theory on $\Phi$ \cite{Banks:2010zn})
is similar to the one of Chern-Simons, but we have two distinct
connections $A_1$ and $A_2$, with $F_2\df dA_2$. It is well know that
whenever $\partial \Phi=0$ this action is gauge invariant (mod 1) and
whenever $\partial \Phi\neq 0$ the action is no longer gauge
invariant. Note that this is precisely the action that arose on the
boundary of our previous example, but we are now interested in taking
this theory and placing it on a manifold with boundary. Our interest
in this action is due to its relation to Maxwell theory: for $k=1$
such an action arises, for instance, as the action for the anomaly
theory for standard $U(1)$ Maxwell theory \cite{Kravec:2013pua,
  Kravec:2014aza, Wang:2018qoy, Seiberg:2018ntt,
  Hsieh:2019iba,Hsieh:2020jpj}, where $A_1$ and $A_2$ are the
background fields for the electric and magnetic 1-form symmetries,
respectively. The higher $k$ cases arise in studying the symmetry
theory for different global forms of the $U(1)$ theory, such as those
remaining on the IR of the Coulomb branch of $\cN=2$ theories
\cite{DelZotto:2022ras}. We will now review how to reformulate this
action more precisely using the language of differential cohomology,
and how to see the non-gauge invariance of the action in this
language.

\medskip

To formulate the $BF$ action we promote $A_1$ and $A_2$ to
differential cohomology cocycles
$\diff a_1, \diff a_2\in \diff Z^{n}(\Phi)$, with $\dim(\Phi)=2n-1$. It is
not difficult to repeat the analysis below for $\diff a_1$ and
$\diff a_2$ of generic degree (as long as they add to one more than
the dimension of $\Phi$), but for the moment we focus on the case where
both differential cohomology classes are of the same degree. The
proper formulation of the $BF$ action in differential cohomology is
then
\[
  \label{eq:k-BF}
  S_{k,BF} = 2\pi i\, k\,\sh\left(\int_\Phi \diff a_1 \cdot \diff a_2\right) \, .
\]
Under gauge transformations $\diff a_i\to \diff a_i+d\diff b_i$, we have
\[
  S_{k,BF} \to S_{k,BF} + 2\pi i\, k\,\sh\left(\int_\Phi d\diff\lambda\right) = S_{k,BF} + 2\pi i\, k\,\sh\left(d\int_\Phi \diff \lambda - \int_{\partial \Phi}\diff \lambda\right)
\]
with
$\diff\lambda \df (-1)^n \diff a_1\cdot \diff b_2+\diff b_1\cdot \diff
a_2+(-1)^nd\diff b_1\cdot \diff b_2\in \diff C(2n)^{2n-1}$ as in the
previous example. We have $R(\diff\lambda)=0$, so
\[
  \sh\left(d\int_\Phi \diff \lambda\right) = \sh(0, -I(\diff\lambda)/\Phi, 0) = -I(\diff\lambda)/\Phi \in \bZ
\]
so if $k\in\bZ$ we can drop the first term of the variation modulo
integers. We end up with
\[
\label{eq:BFvariantion}
  S_{k,BF}\to S_{k,BF} - 2\pi i\, k\int_{\partial \Phi} \sh(\diff \lambda) \mod 1\, .
\]
As promised, the action is gauge invariant (modulo $2\pi i$) on closed
manifolds.

To see the gauge non-invariance on manifolds with boundary, we first
compute
\[
  \sh(\diff \lambda) & = c_1\cup r_2 + (-1)^{n-1} n_1\cup h_2 + r_1\cup \omega_2 + \delta n_1\cup r_2 \\
  & = c_1\cup r_2 + r_1\cup c_2 + \delta n_1\cup r_2 + (-1)^n\left(\sh(d\diff b_1)\cup h_2 + \delta(r_1\cup h_2) \right)\, ,
\]
where in going to the second line we have used $d\diff a_2=\diff 0$. When
integrating over $\partial \Phi$ we can discard the total derivative
term, so we get the gauge variation
\[
  \label{eq:BF-gauge-variation}
  S_{k,BF}\to S_{k,BF} - 2\pi i\, k\int_{\partial \Phi} c_1\cup r_2 + r_1\cup c_2
  +\delta n_1\cup r_2 + (-1)^n\left(\sh(d\diff b_1)\cup h_2
  \right)\mod 1\, .
\]
The fact that this gauge variation does not vanish will be crucial in
the examples we now turn to.

\section{Discrete gauge theories from compactified Maxwell theory}

\label{sec:edge-mode-BF}

A curious (and crucial for us) phenomenon in Maxwell theory, observed
in \cite{Gukov:1998kn,Moore:2004jv,Freed:2006ya,Freed:2006yc}, is that
the operators measuring electric and magnetic flux along torsional
cycles generically do not commute. A consequence of this fact is that
if we place $U(1)$ Maxwell theory on a three-manifold with torsion,
such as the lens space $S^3/\bZ_n$, the resulting one dimensional
theory is non-trivially gapped: a $\bZ_n$ discrete gauge theory
remains after integrating out the massive modes. The topological point
operators that remain are the dimensional reduction of the operators
measuring torsional flux in the $S^3/\bZ_n$ factor. One dimensional
topological field theories are of course rather trivial, but the
techniques we introduce for analysing this case generalise
straightforwardly to higher dimensions, where the answer is more
interesting, so we find this simple example a useful starting point.

Consider the standard $U(1)$ Maxwell theory on
$X\df \bR_t\times (S^3/\bZ_n)$, where we consider $\bR_t$ the time
direction, and we work in the Hamiltonian picture (following
\cite{Freed:2006ya,Freed:2006yc}). Electric and magnetic flux
measuring operators are labelled by elements of $H^1(S^3/\bZ_n;\bT)$,
with $\bT\df\bR/\bZ$. The short exact sequence
\[
  1 \to H^1(S^3/\bZ_n; \bZ)\otimes \bT\to H^1(S^3/\bZ_n; \bT) \to \Tor
  H^2(S^3/\bZ_n; \bZ)\to 1
\]
(from the universal coefficient theorem \cite{Hatcher}) together with
$H^\bullet(S^3/\bZ_n;\bZ)=\{\bZ, 0, \bZ_n, \bZ\}$ implies
$H^1(X;\bT)\cong \Tor H^2(S^3/\bZ_n;\bZ)=\bZ_n$. Accordingly, we
denote the electric flux measuring operators $\Phi_e(\zeta)$ and the
magnetic flux measuring operators $\Phi_m(\xi)$, with
$\zeta,\xi\in H^2(S^3/\bZ_n;\bZ)$. These operators satisfy the
commutation relation~\cite{Freed:2006ya,Freed:2006yc}
\[
  \Phi_e(\zeta)\Phi_m(\xi) = e^{2\pi i\, \sL(\zeta, \xi)} \Phi_m(\xi)\Phi_e(\zeta)
\]
with $\sL\colon H^2(S^3/\bZ_n;\bZ)\times H^2(S^3/\bZ_n;\bZ) \to \bT$
the linking pairing. In the cases of interest to us the lens space
$S^3/\bZ_n$ links a $\bC^2/\bZ_n$ Calabi-Yau singularity, so the
precise $\bZ_n$ action on $S^3$ is generated by
$\rho(z_1,z_2)=(e^{2\pi i/n}z_1,e^{-2\pi i/n}z_2)$. With a suitable
choice of generator $t\in H^2(S^3/\bZ_n;\bZ)$ we then have
$\sL(t,t) = 1/n \mod 1$, see for example
\cite{GarciaEtxebarria:2019caf} for a computation. So, writing
$\zeta=pt$, $\xi=qt$, with $p,q\in \bZ_n$, we have
\[
  \Phi_e(p)\Phi_m(q) = e^{2\pi i pq/n} \Phi_m(q)\Phi_e(p)\, .
\]

When we reduce the theory on $S^3/\bZ_n$ down to one dimension,
because there are no non-zero harmonic 1-forms in $S^3/\bZ_n$, there
are no massless modes arising from the KK reduction, since the mode
constant in the internal space leads to field in 1d which is pure
gauge (since both the characteristic class and curvature automatically
vanish in 1d for degree reasons). Therefore, the one dimensional
theory is gapped and one might be tempted to say that it is
trivial. This is not quite correct: the operators $\Phi_e(p)$ and
$\Phi_m(q)$ remain, with the commutation relation as above. They are
the non-trivial operators in a discrete $\bZ_n$ gauge theory, which
admits a Lagrangian presentation with action
\cite{Banks:2010zn,Kapustin:2014gua}
\[
  \label{eq:1d-gauge-theory}
  S = 2\pi i n \, \sh\left(\int_{M} \diff x \cdot \diff y\right)
\]
where $\diff x, \diff y\in \diff Z^1(M)$ and $M$ is the
one-dimensional manifold where we place the theory. Our goal in the
rest of this section is to reproduce this effective action in two
different ways, both of which prove useful when deriving the SymTFT
from a higher dimensional perspective.

\subsection{Derivation via self-dual formulations of Maxwell theory in four dimensions}\label{sec:Maxwell-reduction}

The main obstruction to using standard ideas about dimensional
reduction is that the degrees of appearing in the one dimensional
action~\eqref{eq:1d-gauge-theory} come from both electric and magnetic
degrees of freedom, so we would need to describe four dimensional
Maxwell theory via an action that includes both electric and magnetic
degrees of freedom. Such a description is in fact available
\cite{Buscher:1987qj,Rocek:1991ps,Schwarz:1993vs,Seiberg:1994rs,Witten:1995gf},
and will lead to the correct result. Although this method will turn
out not to be quite sufficient for our purposes later in the paper, it
illuminates some non-trivial aspects of our later general analysis, so
we will briefly describe it first in the differential cohomology
formulation we are using in this paper.

The construction goes as follows. Standard Maxwell theory in Euclidean
signature is described by an action (we omit the possibility of a
$\theta$ term for simplicity)
\[
  \label{eq:Maxwell-action}
  -S_{g}[\diff a] = \int_{\cM^4} \frac{1}{2g^2} R(\diff a) \wedge
  *R(\diff a)\, ,
\]
with $\diff a \in \diff Z^2(\cM^4)$. In order to formulate this
expression in a way making both electric and magnetic degrees of
freedom manifest we will use, as in \cite{Hsieh:2020jpj}, the
Hopkins-Singer reformulation of the self-dual actions in
\cite{Buscher:1987qj,Rocek:1991ps,Seiberg:1994rs,Witten:1995gf}, also
known in the lattice field theory context as the (modified) Villain
formulation of the $U(1)$ field
\cite{Villain:1974ir,Anosova:2022cjm}. Our presentation is somewhat
unconventional, but it is motivated by the connection to the
Cheeger-Simons formulation later on. In this formulation, we consider
a gauge field $\diff c\in \diff Z^3(\cM^4)$ which is pure gauge
$\diff c=d\diff \ssa$, with $\diff\ssa\in \diff C(3)^2(\cM^4)$. If
$\delta I(\diff \ssa)=0$ then we can reconstruct an
$\diff a\in \diff Z^2(\cM^4)$ by setting\footnote{At this point the
  fact that we are in the lattice is important: in the continuum we
  are taking the curvature to be a differential form, but in the
  lattice it is naturally a real cochain. We elaborate below on the
  relation between cochains and forms.}
\[
  \label{eq:cocycle-from-gauge}
  \diff a(\diff\ssa) \df (I(\diff\ssa), \sh(\diff \ssa), I(\diff\ssa) + \delta \sh(\diff\ssa))\, .
\]
We can enforce the condition that $I(\diff\ssa)$ is closed by
introducing a second gauge transformation parameter
$\diff \ssb\in \diff C(3)^2(\cM^4)$ and modifying the action to be
\[
  \label{eq:Maxwell-self-dual-action}
  -S'[\diff \ssa, \diff \ssb] = -S_g[\diff a(\diff\ssa)] + 2\pi i \sh\left(\int_{\cM^4} d\diff \ssa\cdot \diff \ssb\right) = -S_g[\diff a(\diff\ssa)] - 2\pi i\int_{\cM^4} \delta I(\diff \ssa) \cup \sh(\diff\ssb)\, .
\]
Integration over $\sh(\diff\ssb)$ then implies the desired constraint,
and the theory clearly reduces to standard Maxwell theory. Note that
in this formulation there is no integration over $I(\diff\ssb)$, and
it does not appear anywhere in the action. As explained in detail in
\cite{Anosova:2022cjm} Poisson resummation on $I(\diff\ssa)$
implements S-duality, and leaves us with an equivalent path integral
over $\sh(\diff \ssa)$, $\sh(\diff \ssb)$ and $I(\diff \ssb)$, which
we define to be the dual summation variable that arises in performing
Poisson resummation. The action in the dual path integral is
\[
  -S_m[\diff \ssa, \diff \ssb] = -S_{4\pi^2/g}[\diff b(\diff\ssb)] + 2\pi i\int_{\cM^4} \delta I(\diff \ssb) \cup \sh(\diff\ssa) = -S_{4\pi^2/g}[\diff b(\diff\ssb)] + 2\pi i \sh\left(\int_{\cM^4} d\diff \ssb\cdot \diff \ssa\right)\, ,
\]
where $\diff b(\diff\ssb)$ is defined just as
in~\eqref{eq:cocycle-from-gauge}. If we integrate out $\diff a$ in
this expression we end up with a copy of Maxwell theory with coupling
$4\pi^2/g$ for the field $\diff b\in \diff Z^2(\cM^4)$, which we
interpret as the magnetic dual field.

We now show how to use this viewpoint to
reproduce~\eqref{eq:1d-gauge-theory} by compactification on a lens
space $S^3/\bZ_n$. We restrict to the case in which $\diff \ssa$ and
$\diff \ssb$ are flat, and of the form
$\diff\ssa = \diff \alpha\cdot \diff t$,
$\diff\ssb = \diff \beta\cdot \diff t$, with
$\diff t\in\diff Z^2(S^3/\bZ_n)$ a flat representative of the
torsional generator of $H^2(S^3/\bZ_n;\bZ)=\bZ_n$, and
$\diff \alpha,\diff \beta\in \diff C^0(M)$.\footnote{For notational
  convenience we will leave implicit the pullbacks under the forgetful
  maps $\pi_M\colon M\times S^3/\bZ_n\to M$ and
  $\pi_{S^3/\bZ_n} \colon M\times S^3/\bZ_n\to S^3/\bZ_n$.} Non-flat
choices for the $S^3/\bZ_n$ part of $\diff\ssa$ and $\diff\ssb$ are of
course also present in the four-dimensional path integral, but they
lead to massive modes in the effective one dimensional theory, and we
are only interested in the behaviour at very low energies. We thus
restrict to $\diff t=(t, \varphi, 0)$. Here $t$ is an arbitrary
cocycle representing the generator of $H^2(S^3/\bZ_n;\bZ)$. Since
$\diff t$ is a cocycle, we have $t = -\delta\varphi$. Note that,
despite not explicitly indicating it in the notation, in this equation
we are promoting $t$ to a cochain with $\bR$ coefficients. This
promoted real cochain is trivial in cohomology, which is consistent
since $t$ represents a torsional integral cohomology class.

Clearly $S_g(\diff a) = 0$ due to flatness of $\diff t$, or
equivalently $S_{4\pi^2/g}(\diff b)=0$ in the magnetic formulation, so
the only relevant coupling in the IR is
\[
  S_{d}[\diff \ssa, \diff \ssb] = 2\pi i\sh\left(\int_{\cM^4}d\diff\ssa\cdot\diff\ssb\right)\, .
\]
After a little bit of algebra this gives
\[
  S_d[\diff\ssa, \diff \ssb] = -2\pi i \int_{M\times S^3/\bZ_n} \delta
  \alpha\cup t\cup \beta\cup \varphi = 2\pi i \left(\int_{S^3/\bZ_n} t\cup\varphi\right)\int \alpha\cup\delta \beta\, .
\]
The term in parenthesis is the linking pairing in $S^3/\bZ_n$, giving
(for some convenient choice of generator $t$, other choices lead to
$k^2/n$ mod 1, where $\gcd(k,n)=1$)
\[
\label{eq:linkingp}
  \int_{S^3/\bZ_n} t\cup\varphi = \frac{1}{n} \mod 1\, .
\]
Note also that due to the fact that $\alpha$ and $\beta$ are
multiplied with $t$, which is $n$-torsional, in the path integral we
only need to sum over $\bZ_n$-valued $\alpha$ and $\beta$ cochains, so
the effective path integral in 1d is
\[
  \int [D\alpha][D\beta] \exp\left(\frac{2\pi i}{n} \int_{M} \alpha\cup \delta \beta\right)
\]
with $\alpha,\beta\in C^0(M;\bZ_n)$. This is a well known alternative
presentation of the $BF$ theory~\eqref{eq:1d-gauge-theory}, see for
example appendix B of \cite{Kapustin:2014gua} for a discussion.

\subsection{Derivation via $BF$ theory in five
  dimensions}\label{sec:BF5d}

The previous formulation is quite useful for understanding the physics
of the problem, but it cannot be straightforwardly extended to the
case of a self-dual field, because in this case the initial action
analogous to~\eqref{eq:Maxwell-action} vanishes identically. We now
rederive the results in the previous section from a different
viewpoint, namely $BF$ theory in five dimensions.

In order to do this, first we need to explain how to construct four
dimensional Maxwell theory as an edge mode. We will do so following
(and slightly extending) the ideas
in~\cite{Maldacena:2001ss,Hsieh:2020jpj}.\footnote{We encourage the
  readers to see also \cite{Tong:2022gpg} for an application of these
  ideas to the physics of water.} The beautiful observation in
\cite{Hsieh:2020jpj} was that self-dual fields arise as classical
solutions of Maxwell-Chern-Simons theory exponentially localised on
the boundary. The analysis below presents an essentially
straightforward extension of this argument to Maxwell-$BF$ theory,
which leads to a self-dual formulation of Maxwell theory on the
boundary. Along the way we will elaborate on some aspects that were
left implicit in the analysis in \cite{Hsieh:2020jpj}, in particular
how to relate it to the viewpoint in \cite{Maldacena:2001ss}. Other
important previous works that lead to an understanding of chiral
dynamics from a bulk Chern-Simons perspective include
\cite{Losev:1995cr,Moore:1989yh,Witten:1996hc,Belov:2006jd,
  Belov:2006xj, Costello:2019tri}, where the boundary partition
function is characterised by its properties as a section of an
appropriate line bundle over the space of boundary fields. There are
also many attempts to describe a Lagrangian for chiral fields without
extending to an extra dimension. For example, some early works such as
\cite{Floreanini:1987as,Henneaux:1988gg, Perry:1996mk} construct
Lagrangians by sacrificing Lorentz invariance. There are also works
such as \cite{Pasti:1996vs, Pasti:1997gx, Buratti:2019guq,
  Sen:2019qit, Lambert:2023qgs, Hull:2023dgp, Avetisyan:2022zza} which
manage to construct Lorentz invariant (or covariant) actions through
the introduction of non-polynomial dependence on an auxiliary scalar
field, additional degrees of freedom or extra gauge fields.  We expect
that it should be possible to reproduce our conclusions from these
alternative viewpoints (at least in those cases where they are
developed enough to account for non-trivial topology), as there are
works bridging the approaches \cite{Evnin:2022kqn, Evnin:2023ypu}, but
we have not attempted to do so.

\medskip

Consider the following bulk action on $Y$, with $\dim(Y)=2p+1$ (our main
interest here is $p=2$):
\[
  \label{eq:MCS-BulkAction}
  -S = 2\pi i \int_{Y} \frac{i}{2e^2} R(\diff a_e)\wedge * R(\diff a_e) + \frac{i}{2m^2} R(\diff a_m)\wedge * R(\diff a_m) + k\, \sh(\diff a_e\cdot \diff a_m)
\]
where $\diff a_e, \diff a_m \in \diff Z^{p+1}(Y)$. In the limit
$e^2,m^2\to \infty$ we obtain a $BF$ theory of the type studied above,
while if we set $k=0$ we obtain two decoupled copies of free $p$-form
Maxwell theory in $2p+1$ dimensions.

Let us first work at the level of differential forms (we will extend
our analysis to topologically non-trivial differential cochains
below), by writing $\ssb\df \sh(\diff a_e)$ and
$\ssc\df \sh(\diff a_m)$. 
The holonomies $\sh(\diff a_e)$ and $\sh(\diff a_m)$ are cochains, but
now we wish to build differential forms associated to these
cochains. This is, in general, not a straightforward operation, so let
us briefly elaborate on what we mean, in the context of simplicial
cochains. (We emphasise that we provide the following discussion in
terms of differential forms simply to give a bit more intuition to
those readers unfamiliar with the cochain language; our results are
ultimately expressed fully in the Hopkins-Singer formalism.)

We consider a space $X$ with triangulation $K$. It is perhaps more
natural to recall first the opposite operation, in which we assign
cochains to differential forms. We can straightforwardly assign a
cochain $R\omega\in C^p(X;\bR)$ to every differential form
$\omega\in \Omega^p(X)$ by integration:
\[
  (R\omega)(\Sigma) \df \int_\Sigma \omega
\]
where $\Sigma\in C_p(X)$ is a simplicial $p$-chain. The map
$R\colon \Omega^p(X)\to C^p(Z;\bR)$ is known as the de Rham, or
sometimes ``discretisation'', map.

The Whitney map \cite{Whitney} $W_K\colon C^p(Z;\bR)\to \Omega^p(X)$
is an approximate inverse of $R$: it is defined in such a way that
$RW_K=\mathrm{Id}$, the identity map, and
\[
  \lim_{\text{inf. K}} W_K R = \mathrm{Id}\, ,
\]
where the limit is taken over finer and finer triangulations. We refer
the reader to \cite{Whitney} for the actual definition of the map, and
\cite{Wilson} for further studies of the convergence of the various
cochain constructions to their differential form analogues in the
limit. All we will need is that the map exists, and that it produces
differential forms that capture the behaviour of the smooth real
cochains appearing in the Hopkins-Singer construction in the limit of
infinitely fine triangulations, which is the limit that we will be
implicitly taking when we now switch to the language of differential
forms.

\begin{figure}
  \centering
  \includegraphics[width=0.7\textwidth]{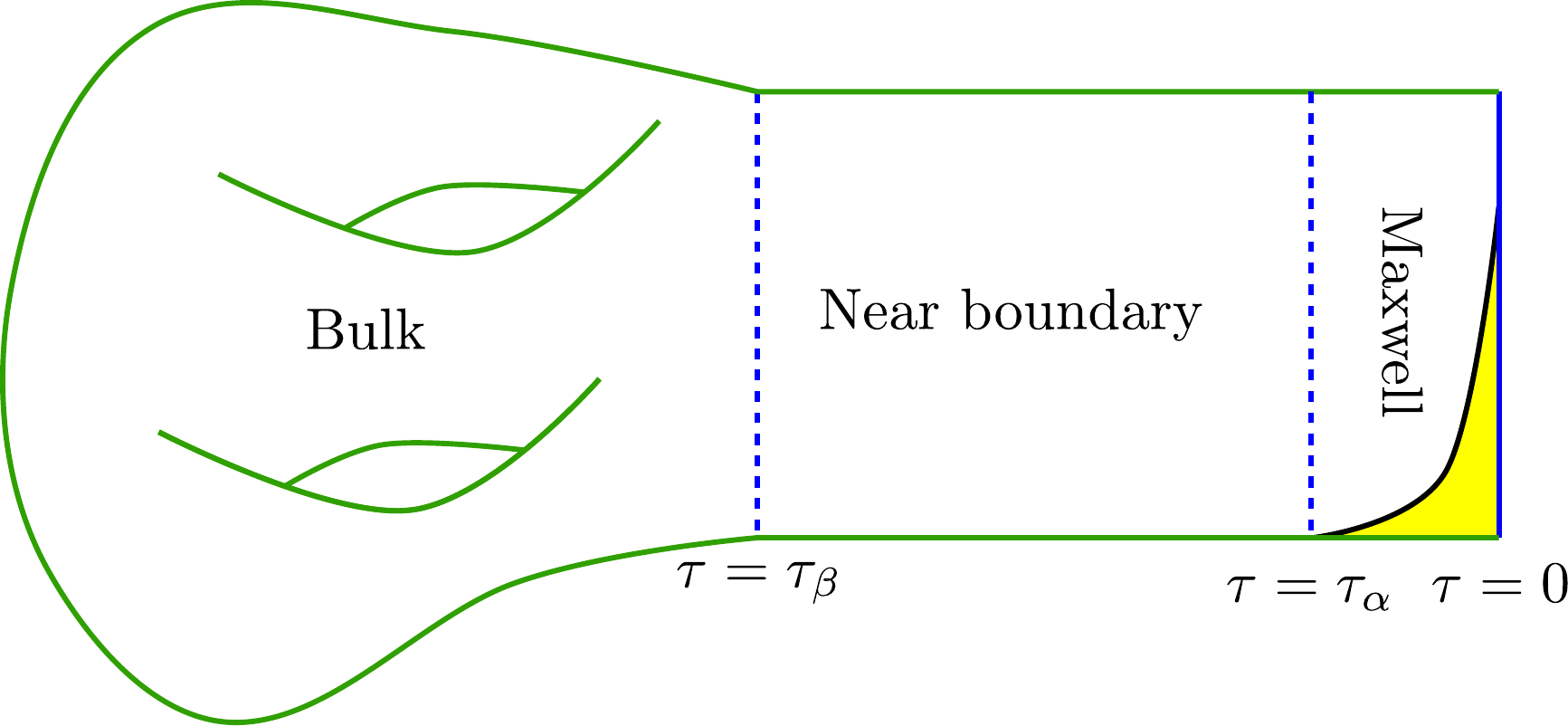}
  \caption{A sketch of the geometry leading to Maxwell theory on the
    boundary. The Maxwell modes are supported on an exponentially
    localised region close to $\tau=0$, which we denote as the
    ``Maxwell'' region. The Maxwell region extends up to
    $\tau=\tau_\alpha<0$, where for $\tau<\tau_\alpha$ we have that
    $e^{\alpha\tau}\ll 1$ (the precise choice of $\tau_\alpha$ is not
    important). The local neighbourhood of the boundary, or ``near
    boundary'' region, is then parameterised by
    $\tau\in[\tau_\beta,\tau_\alpha]$, where $\tau_\beta$
    parameterises where bulk effects become important. We assume
    $|\tau_{\beta}|\gg|\tau_\alpha|$.}
  \label{fig:LocalSolution}
\end{figure}

In terms of these differential forms,~\eqref{eq:MCS-BulkAction}
becomes
\[
  -S = 2\pi i \int_Y \frac{i}{2e^2} d\ssb \wedge * d\ssb + \frac{i}{2m^2} d\ssc\wedge *d\ssc + k \ssb\wedge d\ssc\, .
\]
Under small variations $\ssb\to \ssb+\delta \ssb$ and
$\ssc\to \ssc + \delta \ssc$:
\[
  \label{eq:variation-of-Maxwell-BF}
  -S & \to -S + \int_Y \delta\ssb \wedge \left(\frac{2\pi}{e^2}(-1)^p d*d\ssb + 2\pi i k\, d\ssc\right) + \delta \ssc (-1)^p \left(\frac{2\pi}{m^2} d*d\ssc - 2\pi i k \, d\ssb\right)\\
  & \phantom{\to} - \int_{\partial Y} \frac{2\pi}{e^2}\delta\ssb
  \wedge *d\ssb + \frac{2\pi}{m^2}\delta\ssc\wedge *d\ssc - 2\pi i k\, \delta \ssc\wedge \ssb\, .
\]
The bulk equations of motion are therefore
\begin{subequations}
  \label{eq:MCS-EOMs-local}
  \begin{align}
    \frac{i}{e^2}d* d\ssb - (-1)^p k\, d\ssc & = 0\, ,\\
    \frac{i}{m^2}d* d\ssc + k\, d\ssb & = 0\, .
  \end{align}
\end{subequations}
or in terms of our original differential cochains
\begin{subequations}
  \label{eq:MCS-EOMs}
\begin{align}
  \frac{i}{e^2}d(* R(\diff a_e)) - (-1)^p k R(\diff a_m) & = 0\, ,\\
  \frac{i}{m^2}d(* R(\diff a_m)) + k R(\diff a_e) & = 0\, .
\end{align}
\end{subequations}
These are the equations for massive fields, so far away from the
boundary at $\tau=0$ we expect to recover the $BF$ theory, which is
gapped. But we will now show (slightly extending \cite{Hsieh:2020jpj})
that these equations are also solved by massless fields localised on
the boundary obeying Maxwell's equations. In order to see how these
boundary modes arise, we parameterise a local neighbourhood of the
boundary by $X\times (-\infty,0]$, with the coordinate $\tau$
parameterising a small neighbourhood of the boundary. A sketch of the
geometry is given in figure~\ref{fig:LocalSolution}.

\medskip

Our first task is to choose boundary conditions so that the boundary
terms in~\eqref{eq:variation-of-Maxwell-BF} vanish. We will impose
\[
  \label{eq:bc-local}
  \ssb(\tau=0)=\ssc(\tau=0)=0\, ,
\]
where we are still viewing $\ssb$ and $\ssc$ as differential forms on
$Y$. Our main motivation for imposing this boundary condition is to
connect with the discussion in \cite{Maldacena:2001ss}. We take an
ansatz for the connection near the boundary of the form
\[
  \label{eq:our-MCS-connection}
  \ssb & = (1-e^{\alpha\tau}) F\, ,\\
  \ssc & = (1-e^{\beta\tau}) G\, ,
\]
with $F$ and $G$ forms depending on $X$ only. These solutions do not
vanish away from the boundary, but rather become $F$ and $G$. The
equations of motion set $F$ and $G$ to be flat, which we can interpret
(keeping with our local point of view for the moment, so we can use
the Poincaré lemma) as demanding that $\ssb$ and $\ssc$ are pure gauge
in the near boundary region.

This ansatz is still localised on the boundary in the following way:
via a gauge transformation it can be turned into
\[
  \label{eq:HTY-MCS-connection}
  \ssb = d(e^{\alpha\tau}) \wedge A\, ,
\]
where $F=dA$, and similarly for $\ssc$. These connections do indeed
exponentially localise near the boundary. More meaningfully, the
curvature $d\ssb$ does localise exponentially near the boundary,
assuming $dF=0$:
\[
  d\ssb = -de^{\alpha\tau} \wedge F\, .
\]
The connection~\eqref{eq:HTY-MCS-connection} was in fact the one
proposed in~\cite{Hsieh:2020jpj}; our
ansatz~\eqref{eq:our-MCS-connection} is gauge equivalent but it does
satisfy the boundary condition~\eqref{eq:bc-local}, which is stronger
than the condition $i^*\ssb = i^*\ssc=0$ imposed
in~\cite{Hsieh:2020jpj}, where $i\colon \partial Y\to Y$ is the
inclusion of the boundary into the bulk. We emphasise that the
boundary condition~\eqref{eq:bc-local} is not gauge invariant: this is
desirable since it leads to an interpretation of the boundary modes as
gauge transformations in the near horizon region --- a property
familiar from the holographic description of the Maxwell field given
in \cite{Maldacena:2001ss}. In the semiclassical theory this
identification of $F$ with a gauge parameter also explains the
quantisation of the boundary $U(1)$ connection, which is now
determined by the quantisation of the gauge transformations in the
bulk.

At any rate, our ansatz will satisfy the equations of
motion~\eqref{eq:MCS-EOMs} if
\[
  \label{eq:boundary-Maxwell}
  d(*_X F) & = d(*_X G) = 0\, ,\\
  *_X F & = -i(-1)^p e^2k \frac{\beta}{\alpha^2} e^{\tau(\beta-\alpha)} G\, ,\\
  *_X G & = i m^2 k \frac{\alpha}{\beta^2} e^{\tau(\alpha-\beta)} F\, .
\]
Here $*_X$ denotes Hodge duality on $X$. Consistency with
$*_X(*_XF)=(-1)^p F$ requires $m^2e^2k^2 = \alpha \beta$, and the fact
that the ansatz has $F,G$ constant in the $\tau$ direction requires
$\alpha=\beta$. We see that as long as $\alpha,\beta>0$ the ansatz
localises exponentially on the boundary as we increase $e^2,m^2$, and
the equations of motion for the boundary degrees of freedom are
precisely the Maxwell equations in vacuum, where if we identify $F$
with the electric field strength, $G$ denotes the magnetic field
strength.

\medskip

So far we have been working at the level of differential forms, but it
is important for our purposes to have a differential cochain
description, so that we can describe topologically non-trivial
backgrounds. We do so by promoting the
ansatz~\eqref{eq:our-MCS-connection} to the differential cocycles:
\[
  \diff a_e & = (-\delta f, (1-e^{\alpha\tau})F, -d(e^{\alpha\tau}F))\,,\\
  \diff a_m & = (-\delta g, (1-e^{\alpha\tau})G, -d(e^{\alpha\tau}G))\, .
\]
Here $F\df f+\delta \lambda$, $G\df g+\delta \gamma$, with $f,g$
integral cochains and $\lambda,\gamma$ real ones. We are no longer
imposing that $F$ and $G$ are closed, we allow them to have a
non-closed integral part (which is locally constant, so it does not
affect the analysis above). This is the right global generalisation of
the fact that $F$ should be seen as a gauge parameter in the near
horizon region: neglecting the exponentially decaying terms we have
\[
  \diff a_e & = (-\delta f, f+\delta\lambda, 0)\, ,\\
  \diff a_m & = (-\delta g, g+\delta \gamma, 0)\, ,
\]
which are indeed of the form $\diff a_e=d \diff b_e$,
$\diff a_m=d\diff b_m$ with gauge parameters of the form
$\diff b_e = (-f, -\lambda, 0)$ and $\diff b_m = (-g, -\gamma, 0)$.

\medskip

A summary of the previous analysis is that, after imposing the
boundary conditions~\eqref{eq:bc-local}, the equations of motion of
Maxwell-$BF$ theory for the degree $p+1$ cocycles $\diff a_e$ and
$\diff a_m$ (encoding the information of degree $p$ connections
$\sh(\diff a_e)$ and $\sh(\diff a_m)$) lead to localised modes on the
boundary satisfying Maxwell's equations for $(p-1)$-form gauge fields,
as encoded in~\eqref{eq:boundary-Maxwell}. In the near boundary
region, where the exponentially suppressed curvature is negligible,
the field strength $F$ of the boundary Maxwell theory gets
reinterpreted as a pure gauge $\diff a_e$.

This last observation allows us to connect with the picture in
\cite{Maldacena:2001ss} (also see \cite{Pulmann:2019vrw,
  Avetisyan:2021heg, Avetisyan:2022zza, Fliss:2023dze, Evnin:2023ypu,
  Fliss:2023uiv}). In this picture we send $\tau_\beta$ and
$\tau_\alpha$ to 0, in the notation of
figure~\ref{fig:LocalSolution}. In terms of the couplings of the
Maxwell-$BF$ theory we approach this limit by sending
$e^2,m^2\to\infty$. In the limit we recover pure $BF$ theory in the
bulk. From the point of view of the bulk theory, the boundary
condition is now ``screened'' by the behaviour in the near boundary
region, where the $\diff a_e$, $\diff a_m$ fields become pure
gauge. That is, if we choose to describe the bulk in terms of pure
$BF$ theory, forgetting about the modes localised on the boundary, the
boundary condition we need to impose on the bulk fields as we approach
the boundary is
\[
  \label{eq:BF-BCs}
  [\diff a_e]|_{\partial Y} = [\diff a_m]|_{\partial Y} = \diff 0\, .
\]
Note that here we are only imposing that the differential cohomology
classes of $\diff a_e$ and $\diff a_m$ vanish, not the cocycles
themselves. The gauge transformations that connect different
representatives of the trivial cohomology class are not fixed by this
boundary condition, and as we have seen they furnish the Maxwell
degrees of freedom on the boundary.   When taking the $e^2,m^2\to\infty$ limit, the effect
of the kinetic terms in the bulk gets localised to the
boundary. Following \cite{Maldacena:2001ss}, we have that the dynamics
of the gauge transformations for the pure $BF$ fields in the bulk, or
equivalently of the Maxwell fields on the boundary, are controlled by
an effective boundary Lagrangian
\[
  \label{eq:boundary-term}
  \frac{1}{2e^2} \int_{\partial Y} \ssb\wedge *\ssb\, .
\]
Adding this boundary term to pure $BF$ theory does lead to Maxwell
dynamics on the boundary, as shown in \cite{Maldacena:2001ss},
reproducing the result of the Maxwell-$BF$ analysis above. So this is
the right effective boundary term to add in order to match the
behaviour we found above, but there is one potentially puzzling fact
about~\eqref{eq:boundary-term}: while our original Maxwell-$BF$
action~\eqref{eq:MCS-BulkAction} is symmetric with respect to the
interchange of $\diff a_e$ and $\diff a_m$ (together with $e$ and $m$,
and up to terms coming from the non-commutative behaviour of the cup
products in $\sh(\diff a_e\cdot \diff a_m)$ that we ignore for this
argument), the effective boundary term~\eqref{eq:boundary-term} is not
manifestly symmetric. We emphasise that it does not have to be: all we
should demand of it is that it leads to the right boundary dynamics,
namely Maxwell theory. And in fact, the asymmetry is precisely why we
can write such a simple Lagrangian for modelling the limiting dynamics
on the boundary in the first place! If we tried to
extend~\eqref{eq:boundary-term} to a symmetric Lagrangian of the form
\[
  \label{eq:boundary-term-symmetric}
  \frac{1}{2e^2} \int_{\partial Y} \ssb\wedge *\ssb + \frac{1}{2m^2} \int_{\partial Y} \ssc\wedge *\ssc\, ,
\]
we would effectively be writing a fully democratic action for the
self-dual pair $(F,G)$, the field strength together with its magnetic
dual. An indeed, a short computation where we set $(\ssb,\ssc)=(F,G)$
and use~\eqref{eq:boundary-Maxwell}, shows that the
action~\eqref{eq:boundary-term-symmetric} identically vanishes, just
as the naive action for a self-dual field vanishes.

Finally, note that the boundary action~\eqref{eq:boundary-term} is
clearly not gauge invariant, a remnant of the fact that our original
boundary conditions~\eqref{eq:bc-local} were not gauge invariant
either. In the discussion below we find it useful to work in this
infinitely massive limit, so that the bulk is pure $BF$ theory.

\medskip

This might all sound somewhat exotic, but a context which might be
more familiar were something similar happens (with ``boundary''
replaced by ``brane'') is D-brane physics, where the
``gauge-invariant'' field strength on a single D-brane is given by
$F-B$, with $B$ the bulk NSNS two-form field. In this case we treat
$F$ as a dynamical field, and $B$ as the restriction of the bulk
field, subject to the gauge identifications $F\to F+\Lambda$,
$B\to B - \Lambda$ with $\Lambda$ an integrally quantised differential
form of degree two on the worldvolume of the brane. (Small gauge
transformations have $\Lambda=d\lambda$, with $\lambda$ a 1-form
connection on a higher $U(1)$ bundle.) Thanks to these gauge
transformations, we can trade off any choice of $F$ by a choice of
gauge representative of $B$. The Maxwell action in this gauge is
precisely~\eqref{eq:boundary-term}. This bulk viewpoint on brane
degrees of freedom can sometimes be very useful, see for instance
\cite{Douglas:2014ywa}.

This point of view raises a potentially interesting connection with
recent work on non-invertible symmetries realised in string theory,
which we will only sketch: we have just argued that the worldvolume
theory on (abelian) branes can be understood in terms of the bulk
gauge transformations becoming physical. In the case of finite
symmetries, making a gauge field physical on a submanifold, or
equivalently ``ungauging it'', is equivalent to gauging its dual field
on the submanifold. Gauging discrete fields on submanifolds is
precisely the operation known as ``higher gauging''
\cite{Roumpedakis:2022aik} or ``condensation''
\cite{Gaiotto:2019xmp,Choi:2022zal}. So the discussion above suggests
that branes should encode condensation defects, at least in some
cases. This is indeed the case, as has been understood recently
\cite{Apruzzi:2022rei, GarciaEtxebarria:2022vzq, Heckman:2022muc,
  Heckman:2022xgu, Etheredge:2023ler, Dierigl:2023jdp,Bah:2023ymy,
  Apruzzi:2023uma, Yu:2023nyn}.

It is a natural question whether this understanding of branes in terms
of ungaugings can be made fully precise, and how far it goes. We will
not attempt to push it further in this paper, and only mention that a
very related idea has been recently advocated by Donagi and Wijnholt
for the case of non-abelian stacks in M-theory \cite{Donagi:2023sbk}.

\paragraph{Back to the SymTFT.} We now have a description of Maxwell
theory in terms of $BF$ theory on the bulk. Crucially, this
description treats electric and magnetic degrees of freedom on an
equal footing (up to a subtlety we will mention momentarily), so it is
a good candidate to make the emergence of non-commutativity between
electric and magnetic fluxes on the boundary manifest. As we argued
above, after dimensional reduction on the lens space $S^3/\bZ_n$ this
non-commutativity effect should lead to a one-dimensional theory on
the boundary with action
\[
  S = 2\pi i n\, \sh\left( \int_{M} \diff x\cdot \diff y\right)
\]
with $M$ the boundary manifold. This result follows from what we have
derived so far: we have identified the gauge fields on the boundary
with the gauge transformations $\diff b_i$ at the boundary, so to
understand the effective theory that we obtain after reducing the
theory on the boundary on $S^3/\bZ_n$ we look at how the $BF$ action
on a space with boundary changes under gauge transformations of the
form $\diff a_i\to \diff a_i+d\diff b_i$, which was derived
in~\eqref{eq:BF-gauge-variation} above. Our boundary
conditions~\eqref{eq:BF-BCs} require that $\diff a_e$ and $\diff a_m$
are always pure gauge near the boundary, so we might as well set the
initial values to zero in~\eqref{eq:BF-gauge-variation} when studying
the effective action for the gauge transformations. We obtain
\[
\label{eq:btMaxwell}
  S_{\text{bulk}} \to S_{\text{bulk}} + 2\pi i\int_{\partial Y} \delta n_1 \cup r_2
\]
which is precisely the term appearing in the ``almost democratic''
action~\eqref{eq:Maxwell-self-dual-action} for Maxwell theory (in
slightly different notation). The rest of the derivation now proceeds
identically to the discussion in the previous section.  Note that from
this point of view the perhaps somewhat unfamiliar ``Villain''
characterisation of gauge fields in Maxwell theory as elements of
$C(3)^2(\cM^4)$ becomes perfectly natural, as this is precisely the
group where gauge transformations of the bulk fields live.

\medskip

Note that in our discussion, it was natural to restrict to those gauge
transformations on the boundary that are constant in the integration
fibre. Since the change in the effective action depends only on the behaviour of the gauge transformation on the boundary, we are free to
choose the gauge transformation in the bulk as we wish. A convenient choice is to restrict ourselves to bulks that preserve the fibration
structure of the boundary, and to gauge transformations that are
constant along the fibre also in such bulks. If we make these choices,
then we can first integrate over the fibre in the bulk, and then study
the effect of induced gauge transformations on the resulting
theory. It is in this way that figure~\ref{fig:approach-summary} can
be made precise. Denoting by $\diff\Omega$ the Chern-Simons
differential cocycle in the bulk with gauge transformation
$\diff \Omega \to \diff \Omega + \Delta \diff \Omega$, and
$\cL_\SymTFT$ the Lagrangian (understood as a real cochain) for the
SymTFT arising after integration of the edge mode theory on $L$, a
precise mathematical formulation of figure~\ref{fig:approach-summary}
is therefore
\[
  \label{eq:symm-inflow}
  \sh\left(\int_L \Delta\diff\Omega\right) = \delta \cL_\SymTFT\, .
\]
This equation is quite reminiscent of the kind of equation that
appears when doing anomaly descent, but now it includes information
not only about the anomalies, but rather the full SymTFT. We will
expand on this point below.

As an example, let us go back to our working example of the $BF$
theory in five dimensions, with $L=S^3/\bZ_n$. The differential cocyle
representing the five dimensional $BF$ theory (once we take its
holonomy) is
\[
  \diff \Omega = \diff a_e\cdot \diff a_m\, ,
\]
with $\diff a_e,\diff a_m\in \diff Z^3(Y)$. If we write
$\diff a_i=\diff \alpha_i\cdot \diff t$, with
$\diff \alpha_i \in \diff C^1(\cB)$ (with $\cB$ a bulk such that
$\partial \cB=M$) and $\diff t$ a flat representative of
$\diff Z^2(L)$ as above, we find
\[
  \int_{S^3/\bZ_n} \diff\Omega = (0, \ell\, I(\diff \alpha_e) \cup I(\diff \alpha_m), 0) \mod 1\, .
\]
Here $\ell\df\int_{S^3/\bZ_n}\sh(\diff t\cdot\diff t)$ is rational
number that agrees modulo one with the torsional linking pairing of
$I(\diff t)$ with itself. For instance, if we choose $\diff t$ a flat
uplift of a suitable generator of $H^2(S^3/\bZ_n;\bZ)$, we have
$\ell=1/n$ mod 1. Now consider the case in which $\diff a_i$ is pure
gauge, namely $\diff a_i= d\diff\lambda_i$, with
$\diff\lambda_i=\diff\beta_i\cdot \diff t$. Taking $\diff a_i$ to be
a pure gauge is equivalent to first performing the gauge
transformation $\diff a_i= \diff a_i+d\diff\lambda_i$ and so
$\diff \Omega \to \diff \Omega + \Delta \diff\Omega$, and then setting
$\diff a_i=\diff 0$ to be left with
$\Delta \Omega=d\diff\lambda_e\cdot d\diff\lambda_m$. Therefore we can
write
\[
   \sh\left(\int_{S^3/\bZ_n} \Delta\diff \Omega\right) = \delta\bigl(\ell \beta_e\cup \delta \beta_m\bigr)\, ,
\]
and we conclude that (as expected)
$\cL_\SymTFT=\ell \beta_e\cup \delta \beta_m$.

\subsection{Inclusion of backgrounds}

Going back to 4 dimensions, recall that in section \ref{sec:BF5d}, we
have imposed boundary conditions
$[\diff a_e]|_{\tau=0} = [\diff a_m]|_{\tau=0} = \diff 0$, which lead
to ordinary Maxwell theory on the boundary.

More generally, we can couple Maxwell theory to currents for the
electric and magnetic 1-form symmetries. We do this by imposing
\[ [\diff a_e]|_{\tau=0} = [\diff j_e] \qquad ; \qquad [\diff
  a_m]|_{\tau=0} = [\diff j_m]\, ,
\]
for fixed differential cohomology classes $[\diff j_e], [\diff j_m]$
satisfying $I([\diff j_e]) = I([\diff j_m]) =0$ (so that the partition
function does not vanish \cite{Hsieh:2020jpj}). If we pick
representative cocycles $\diff j_e$, $\diff j_m$ for $[\diff j_e]$,
$[\diff j_m]$, the sum over gauge transformations of $\diff a_i$ left
unfixed by the boundary condition becomes the path integral for
Maxwell theory in the presence of background currents $\diff j_e$,
$\diff j_m$ for the 1-form symmetries of Maxwell. (Different choices
of representatives amount to a harmless redefinition of the Maxwell
fields in the path integral.) We refer the reader to
\cite{Freed:2000ta} for an illuminating discussion of Maxwell theory
in this formalism.

In detail, this goes as follows. Since $I([\diff a_i])|_{\tau=0}=0$,
up to gauge transformations we can represent $\diff a_i$ by
differential forms $B_i\in \Omega^2(X)$, so that
$\diff a_i = (0, B_i, dB_i)$ \cite{Hsieh:2020jpj}.
So, up to gauge transformations, there exist cochains
$\diff B_i \df (0,0, B_i)$ such that
$\diff a_i|_{\tau=0}=d\diff B_i$. We interpret the differential forms
$B_i$ as the background fields for the higher form
symmetry. The fields $B_i$ are defined only up to gauge
transformations, and in particular the quotient by large gauge
transformations makes the physical information live in
$ \Omega^2(X)/\Omega^2_\bZ(X)$, where $\Omega^2_\bZ(X)$ indicates the
integrally quantised differential forms on $X$.

Putting the gauge transformations back into place, when doing the path
integral described above we end up with an action where the gauge
parameters $\diff F,\diff F^D\in C(3)^2(X)$ appearing the original
Maxwell kinetic term now appear shifted by the background fields. It
is customary to denote these shifted fields by
$\diff \cF\df \diff F - \diff B_e$,
$\diff \cF^D\df \diff F^D - \diff B_m$.

We emphasise that $\diff \cF, \diff \cF^D$ are not gauge parameters in
general. (Although the difference of two such fields is a gauge
parameter.) In other words, they cannot always be interpreted as
differential cocycles describing an ordinary $U(1)$ bundle. As a
simple example, if we want to couple the Maxwell theory on the
boundary to a \emph{flat} (in addition to topologically trivial)
background electric current $B^e\in \Omega_{\text{closed}}^2(X)$ and
no magnetic current, we can take $\diff j_e = d(0,0, B^e)=(0, B^e,0)$
and $\diff j_m=\diff 0$. For generic choices of $B^e$ we have that
$\diff j_e$ is not pure gauge (since the holonomies of a pure gauge
field are integrally valued, while the holonomies of $B^e$ are not).

There is of course no requirement that the backgrounds $B_i$ are flat,
only that they are topologically trivial. Whenever $dB_i\neq 0$, the
bulk theory that we have described can be understood as a dynamical
version of the familiar anomaly theory for the $U(1)^e\times U(1)^m$
1-form symmetry in Maxwell theory.

\section{Example: the $BF$ sector of the SymTFT for the 6d $(1,1)$
  theory}
\label{sec:(1,1)-example}

As an example of the previous discussion, we rework from our viewpoint
the $BF$ sector of the $(1,1)$ theories in six dimensions
\cite{GarciaEtxebarria:2019caf}, describing the structure of discrete
1-form and 3-form symmetries of the theory. The 2-form symmetries in
6d $(2,0)$ and $(1,0)$ theories
\cite{Tachikawa:2013hya,Monnier:2017klz, DelZotto:2015isa} can be
analysed similarly, by studying the reduction of a Chern-Simons theory
with boundary mode the self-dual $F_5$ field in type IIB string
theory. For the 6d $(1,1)$ theory constructed by putting IIA string
theory on $\bC^2/\Gamma$, with $\Gamma$ a discrete subgroup of
$SU(2)$, the relevant $BF$ theory is a discrete gauge theory for the
group $\Gamma^{\text{ab}}\df \Gamma/[\Gamma,\Gamma]$, the
abelianisation of $\Gamma$. For instance, for $\Gamma=\bZ_n$ we have
$\Gamma^{\text{ab}} = \bZ_n$. (The rest of the cases are listed in
many references, see for example \cite{DelZotto:2015isa}.) In the
$\Gamma=\bZ_n$ case we have
\[
\label{eq:7dSymTFT}
  S_{\SymTFT}[X^7] = \frac{2\pi i}{n}\int_{X^7} c_2\cup \delta c_4
\]
with $c_i\in C^i(X^7; \bZ_n)$ for $i=2,4$. In what follows we will
assume that $X^7$ has no torsion, for simplicity. From the arguments
in \cite{GarciaEtxebarria:2019caf} it is clear that the background
fields $c_2$ and $c_4$ arise respectively from $F_4$ and $F_6$ fields
in IIA on $S^3/\bZ_n\times X^7$. A full treatment requires that we
think of these fields in terms of $K$-theory. We will make more comments about this point in section~\ref{sec:couplings}.

Just as in previous examples, the type IIA fields can arise as
boundary gauge degrees of freedom of bulk fields in 11d
on $S^3/\bZ_n\times Y^8$ with boundary
$\partial (S^3/\bZ_n\times Y^8)=S^3/\bZ_n\times X^7$. So
starting with the bulk fields as differential cocycles
$\diff a_j\in \diff Z^j(S^3/\bZ_n\times Y^8)$ for $j=i+3=5,7$, the
action governing the dynamics of these cocycles is
\[
    \label{eq:11dBF}
  S_{BF} = 2\pi i\, \sh\left(\int \diff a_5\cdot \diff a_7\right)\, .
\]
To see the boundary modes, we follow an analogous discussion to
section \ref{sec:BF5d}. That is we impose the boundary condition
\begin{equation}\label{eq:BC11d}
    [\diff a_j]\mid_{\partial Y}=\diff 0
\end{equation}
and study the dependence of the action on the gauge transformations.

The 7d symmetry theory results from the gauge non-invariance of the
action (\ref{eq:11dBF}) on the boundary after dimensional reduction on
$S^3/\bZ_n$. Explicitly, we expect:
\[
\label{eq:Dela5a7}
  2\pi i\, \sh\left(\int_{S^3/\bZ_n} \Delta (\diff a_5\cdot \diff a_7)\right)= \delta \cL_{\SymTFT}\, ,
\]
under the condition that $\diff a_j$ are pure gauge on the boundary.
Similarly to (\ref{eq:transaa}), we find that under the gauge transformations $\diff a_j \to \diff a_j+d\diff b_{j-1}$,
\[
    \diff a_5\cdot \diff a_7 \to \diff a_5\cdot \diff a_7 + \Delta (\diff a_5\cdot \diff a_7)= \diff a_5\cdot \diff a_7 +
    d\diff \lambda\, ,
\]
with
\[
    \diff \lambda= 
    (-1)^{|\diff a_5|} \diff a_5\cdot \diff b_6+\diff b_4\cdot \diff a_7+ (-1)^{|\diff a_5|}d\diff b_4\cdot \diff b_6\, .
\]
We may expand
$\diff b_{4}= \diff \beta_2\cdot \diff t$ and
$\diff b_{6}= \diff \beta_4\cdot \diff t$, with
$\diff \beta_{i}=(\tilde{c}_{i},r_i,0)$ differential cochains on $X^7$ extending to $Y^8$, and $\diff t=(t, \varphi, 0)$ as in the example of section \ref{sec:Maxwell-reduction}.\footnote{As in the previous examples, we leave implicit the pullbacks under the projections to $S^3/\bZ_n$ and $Y^8$.} Then, substituting these in the left hand side of
(\ref{eq:Dela5a7}) we find
\[
    \sh\left(\int_{S^3/\bZ_n} \Delta (\diff a_5\cdot \diff a_7)\right)
    &=
    \sh\left(\int_{S^3/\bZ_n}  d\diff b_4\cdot d\diff b_6\right)\mod 1\\
    &=
    \sh\left(\int_{S^3/\bZ_n} \diff t\cdot \diff t\right)
    I( d\diff \beta_2\cdot d\diff \beta_4 )
    \mod 1\\
    &= \delta\, \biggl(  \frac{1}{n}\, \tilde{c}_{2} \cup \delta \tilde{c}_{4} \biggr) \mod 1
    \, ,
\]
using (\ref{eq:linkingp}).
To see the consistency with (\ref{eq:7dSymTFT}), note that the fields 
$c_i\in C^i(X^7; \bZ_n)$ and $\tilde c_i\in C^i(X^7; \bZ)$ 
are related by $c_i=\tilde{c}_i \mod n$, and so 
\[\frac{ c_2 \cup \delta c_4}{n} =\frac{ \tilde{c}_{2} \cup \delta \tilde{c}_{4}}{n} \mod 1\, .\]
Alternatively, we may represent this action in terms of $U(1)$ fields
$c_i^{U(1)}\in C^i(X^7; U(1))$, where $c_i^{U(1)} = c_{i}/n$,
and write
\[
  S_{\SymTFT}[X^7] = 2\pi in\int_{X^7} c^{U(1)}_2\cup \delta c^{U(1)}_4\, .
\]

\section{Quadratic refinements}
\label{sec:quadratic-refinements}

So far we have been somewhat cavalier about the factor of $k/2$ in
front of the Chern-Simons term~\eqref{eq:CS-heuristic}. The
Chern-Simons theory relevant for describing the self-dual fields in string theory is the one at $k=1$, so we need to slightly
refine the discussion above. This can be done in terms of
\emph{quadratic refinements}. Consider the symmetric bilinear pairing
\[
  B_k(\diff a, \diff b) = k\, \sh\left( \int_Y \diff a\cdot \diff b\right) \mod 1 \in \diff
  H^1(\pt) = \bR/\bZ
\]
with $\diff a, \diff b\in \diff H^{2n+2}(Y)$ and $\dim(Y)=4n+3$. We
say that $q_k\colon \diff H^{2n+2}(Y)\to \bR/\bZ$ is a quadratic
refinement of $B_k$ if
\[
  B_k(\diff a, \diff b) = q_k(\diff a + \diff b) - q_k(\diff a) -
  q_k(\diff b) + q_k(\diff 0)\, .
\]
We include the $q_k(\diff 0)$ term since it can be important when dealing
with the purely gravitational sector, but in our examples below we can
take $q_k(\diff 0)= 0$. It is clear that whenever $k$ is even
$q_k^{\text{even } k}(\diff a)\df \frac{k}{2}\, \sh\left(\int_Y\diff a\cdot\diff a\right)$
is indeed a quadratic refinement of $B_k$, but the quadratic
refinement is more fundamental, as it also makes sense for odd $k$.

A way of constructing quadratic refinements was given in
\cite{Witten:1996hc,Freed:2000ta,Hopkins:2002rd}, which we now briefly
summarise. We start by introducing a differential Wu cocycle
$\diff \lambda=(c,h,w) \in \diff Z^{2n+2}(Y)$ which is such that
$c=\nu_{2n+2} \mod 2$, with $[\nu_{2n+2}]$ the degree-$(2n+2)$ Wu
class. (In the special case of $n=0$ a choice of Wu cocycle is
equivalent to a choice of spin structure \cite{Hopkins:2002rd}.) We
then define the quadratic refinement of the action to be
\cite{Hopkins:2002rd,Belov:2006jd}\footnote{In fact, we need to add
  $\frac{k}{8} \int (\diff \lambda\cdot \diff \lambda - \diff L)\, $
  to~\eqref{eq:cswu1}, where $L$ is the Hirzebruch polynomial
  \cite{Hopkins:2002rd}. Such a term will not affect the simple
  examples we discuss, so we omit it for conciseness, but it will be
  required in the analysis of more complicated examples.}
\begin{equation}\label{eq:cswu1}
  \mathsf{CS}_{\diff\lambda}[\diff a]= \frac{k}{2}\, \sh\left(  \int_Y  \diff a \cdot (\diff a-\diff \lambda) \right)\mod 1\, .
\end{equation}
The case of most interest to us will be $4n+3=11$. We assume that we
want to define our Chern-Simons term on $L\times M$, where $L$ is the
internal (closed) compactification space. If we assume that $M$ is
also closed, then the manifold $L\times M$ admits a spin extension to
12 dimensions (since the 11 dimensional spin bordism group
vanishes). In this case we can view the 11d Wu class
$[I(\diff \lambda_{6})]$ as a restriction of the 12-dimensional Wu
class. Luckily, the 12-dimensional Wu class is zero mod 2 for Spin
manifolds, as pointed out in \cite{Witten:1996hc}.\footnote{The Adem
  relation $\Sq1 \Sq4=\Sq5$ and $\Sq2 \Sq4=\Sq6+ \Sq5 \Sq1$ give
  $\Sq6=\Sq2 \Sq4+\Sq1 \Sq4 \Sq1$. So for any $x\in H^6(Y;\bZ_2)$
  \[
    \nu_6\cup x=\Sq6(x)=\Sq2 (\Sq4(x))+\Sq1 (\Sq4(\Sq1(x)))\, ,
  \]
  which vanishes on Spin manifolds. This is because $\Sq2$ and $\Sq1$,
  as maps to the top-dimensional degree, are given by multiplication
  by the second and first Wu classes, respectively, and these both
  vanish on Spin manifolds.} Thus, its integral lift and its
restriction to 11d can be taken to be trivial:
$[I(\diff \lambda_{6})]=0\mod 2$. So in 11 dimensions, we may choose
$\diff \lambda$ to be 0, or more generally twice some differential
cocycle $\diff C$. The case of interest to us is when
$\partial M\neq 0$, so the discussion above needs to be modified to
incorporate manifolds with corners. We will not attempt to do this in
this paper, although given that the reasoning above is mostly local
(apart from the bordism argument), it seems natural to conjecture that
$\diff \lambda = \diff 0$ is also a valid choice in this case.

The dependence of the Chern-Simons term on the differential
Wu cocycle is given by the formula \cite{Hopkins:2002rd}
\begin{equation}
  \mathsf{CS}_{\diff\lambda-2\diff b}[\diff a]=\mathsf{CS}_{\diff\lambda}[\diff a+\diff b]\, .
\end{equation}
Thus, for $\diff \lambda=\diff 0$ we have (from~\eqref{eq:cswu1})
\begin{equation}
    \mathsf{CS}_{-2\diff b}[\diff a]=\mathsf{CS}[\diff a+\diff b]
    = k\, \sh\left(\int_{Y} \diff a\cdot\diff b\right) 
    +\mathsf{CS}[\diff a]+\mathsf{CS}[\diff b]\mod 1
    \, .
\end{equation}
The last equality is the statement that the Chern-Simons is a
quadratic refinement of the bilinear pairing
$ B(\diff a,\diff b)= k\, \sh\left(\int_{Y} \diff a\cdot\diff
  b\right)\mod 1$.

\section{Couplings}

\label{sec:couplings}

Our goal in this paper is to give a unified prescription for deriving
the SymTFT, but so far we have only discussed how to derive the $BF$
sector (or more generally, abelian Chern-Simons sectors). We will now
present some evidence that suggests that in favourable cases the
anomalies can also be naturally incorporated if one represents the
type II RR fields as $H$-twisted differential $K$-theory cocycles, and
generalises the bulk Chern-Simons theory they come from to an
$H$-twisted differential K-theory version of Chern-Simons theory. That
this works is not surprising: in \cite{Apruzzi:2021nmk} the anomaly
terms appeared from integrating certain topological terms in the
string theory action on the horizon of the singularity, and the
analysis of Belov and Moore \cite{Belov:2006xj} shows that the
relevant terms in the string theory action appear from the
differential $K$-theory version of the Chern-Simons action in eleven
dimensions. We will argue in particular that our inflow
prescription~\eqref{eq:symm-inflow} also reproduces the anomaly
terms.

Our analysis in this section is preliminary in two important
respects. First, we will be approximating differential K-theory by
ordinary differential cohomology, keeping track of the extra couplings
induced in the Chern-Simons action by the $H$-twisting. This is a
technical restriction in order to avoid introducing additional
technology, but it would certainly be interesting to do a proper
differential K-theory treatment and remove this approximation. More
importantly, we can only derive anomaly terms in this way if we know
the right geometric formulation of the ``bulk'' Chern-Simons theory
with our desired modes as edge modes. So at this moment we cannot do
M-theory,\footnote{Following \cite{Fiorenza:2019usl}, a candidate for
  extending our approach to M-theory would be to study $J$-twisted
  cohomotopy Chern-Simons on manifolds with boundary.} or
configurations that require a formulation where $H$ is also dynamical
(for example if both the electric and magnetic NSNS fields, $B_2$ and
$B_6$, play roles in the analysis), as $H$-twisted $K$-theory seems to
be no longer a good approximation. We refer the reader to
\cite{Evslin:2006cj} for a discussion of the puzzles that arise when
trying to model this more general situation.

We emphasise that our approach and the standard anomaly inflow picture
deal with anomalies very differently: in standard anomaly inflow, one
has a Chern-Simons action $S_{\text{anomaly}}(\diff B)$ in
$(d+1)$-dimensions, where the $\diff B$ are background fields for the
symmetries of the $d$-dimensional theory. There is also a
$(d+2)$-dimensional action $S_{I}(\diff B)$, the anomaly polynomial,
given by $S_I(\diff B)=\delta S_{\text{anomaly}}(\diff B)$. In all
these actions the fields $\diff B$ are classical. On the other hand,
in our context we have the SymTFT $S_{\SymTFT}(\diff\ssb)$, depending
on dynamical fields $\diff\ssb$, and the $(d+2)$-dimensional theory is
also a dynamical (and conjecturally K-theoretical) Chern-Simons theory
$S_{\text{inflow}}(\diff a)$ for dynamical fields $\diff a$. As we
have argued above, the relation between $\diff a$ and $\diff\ssb$ is
that $\diff\ssb$ are the gauge parameters for $\diff a$, and
$\Delta S_{\text{inflow}}(\diff a)=\delta S_{\SymTFT}(\diff\ssb)$.

\medskip

As an example, consider IIA string theory on a $\bC^2/\bZ_n$
singularity, as in section~\ref{sec:(1,1)-example}. If we choose the
$SU(n)$ global form this theory has a $\bZ_n$ 1-form symmetry,
measuring the $n$-ality of Wilson line insertions. In the string
theory construction, the Wilson lines arise from D2 branes wrapping
non-compact 2-cycles on the $\bC^2/\bZ_n$ geometry, and their
remaining direction is the worldline of the Wilson line in the field
theory. We, therefore, identify the background field for the $\bZ_n$
1-form symmetry of this theory with $\diff F_4$
\cite{GarciaEtxebarria:2019caf}. There is also an instanton $U(1)$
1-form symmetry. The background for this symmetry is a degree 3
cocycle $\diff G$, with connection $B$. There is a mixed anomaly
between the 1-form $\bZ_n$ symmetry and the 1-form $U(1)$ instanton
symmetry, of the schematic form
\[
\label{eq:1form-instAnom}
  S_{\text{anomaly}} = 2\pi i\int_{X^7} dB \cup n_{\text{inst}}(C_2)\, ,
\]
where $C_2\in Z^2(X^7;\bZ_n)$ is a bulk extension of the background
for the 1-form symmetry, and $n_{\text{inst}}(C_2)\in\bQ/\bZ$ is the
fractional part of the instanton number of the gauge bundle in the
presence of the 1-form background $C_2$. We will not need its precise
expression, but it can be found for example in
\cite{Witten:2000nv,Apruzzi:2020zot}.  We will now show that there is
a similar term in the SymTFT for the system, with $C_2$ now taken to
be a dynamical $\bZ_n$-valued 2-cochain denoted by $c_2$. The standard
anomaly is then reproduced when we take Dirichlet boundary conditions
for $c_2$ on the gapped boundary of the SymTFT. We will keep $B$,
coming from the NSNS 2-form field $B_2$, as a classical field
throughout, due to the limitation of the formalism mentioned above.

\medskip

This term can indeed be argued to be present as follows. The RR fields
in IIA string theory can be modelled in terms of a self-dual RR field
$\diff F\in \diff K^{-1,\diff H}(\cM)$ in $\diff H$-twisted
differential $K$-theory \cite{Moore:1999gb},\footnote{ For material on
  differential $K$-theory, we refer the reader to \cite{Freed:2009mw},
  and \cite{Freed:2006yc, Hsieh:2020jpj} as well as the closely
  related formulations in \cite{lott_1994, simons2008structured,
    tradler2012elementary}. Alternative descriptions of differential
  $K$-theory are given in \cite{Hopkins:2002rd,bunke2009smooth,
    gorokhovsky2018hilbert}.} where $\diff H\in \diff H^3(\cM)$ is the
NSNS field with field strength $H\df R(\diff H)$. Similar to the
differential cohomology construction we had before, we may realise
$\diff F$ as the boundary mode of an 11-dimensional Chern-Simons
action on $\cN$ for a field which is an element of the differential
$K$-theory group $\diff K^{0,\diff H}(\cN)$, with $\partial \cN=
\cM$. We start by considering the topologically trivial case, so we
take $c\in \Omega^{\text{even}}(\cN)$ to be
\[
c=c_0+c_2+c_4+c_6+c_8+c_{10}\, ,
\]
a sum of differential forms of even degree, and define $d_H=d-H$ to be
the DeRham differential twisted by $H$. Then, following the work of
Belov and Moore \cite{Belov:2006jd,Belov:2006xj} (to which we refer
the reader for more details, see also \cite{Freed:2000ta,
  Hopkins:2002rd}), we take the K-theoretical Chern-Simons action to
be
\begin{equation}\label{eq:CdHC}
    \mathsf{CS}_{H}[c] = \frac{1}{2}\int_\cN c\wedge (d_{H^*}c^*)\, ,
\end{equation}
where $c_{2k}^*=(-1)^kc_{2k}$ and $H^*=-H$. One may reproduce the
supergravity equations by adding a kinetic term to this Chern-Simons
and substituting a suitable ansatz into the equation of motion.

It is important to note at this point that the full Chern-Simons is
given by a quadratic refinement, as in
section~\ref{sec:quadratic-refinements}, that includes several other
terms.\footnote{See for instance \cite{Hsieh:2020jpj} for concrete
  expressions for this quadratic refinement in differential
  $K$-theory.} In particular, it includes an eta invariant term which
in differential $K$-theory is the only term that encodes information
about the topological sector. As we have seen it is this topological
data that gives rise to a non-trivial SymTFT for the finite group
symmetries. Thus, to do the computation in twisted differential
$K$-theory, we must understand how the eta invariant transforms under
the gauge transformation of the Chern-Simons field, and then we may
apply the same general formula \eqref{eq:symm-inflow}. However, to
simplify the analysis we do not do this, and we will instead recover
some of the topological data by refining~\eqref{eq:CdHC} to ordinary
differential cohomology. As we will see, doing so correctly reproduces
the anomalies which we have seen to arise from the string theory
topological Chern-Simons action as studied in \cite{Apruzzi:2021nmk},
which supports our expectation that the proper K-theory formulation
will also give the right result.

\medskip

To refine the Chern-Simons (\ref{eq:CdHC}) to differential cohomology,
let us first rewrite it as
\begin{equation}
  \label{eq:CS_bulk_expanded}
  \mathsf{CS}_{H}[c] = \frac{1}{2}\int_\cN c\wedge dc^*+ c\wedge H\wedge c^*\, .
\end{equation}
It is easy to see, at least at the level of differential forms, that
the second term in~\eqref{eq:CS_bulk_expanded} gives the kind of
$C\wedge dC\wedge H$ Wess-Zumino term that according to the
prescription
in~\cite{Bah:2019rgq,Bah:2020jas,Cvetic:2021sxm,Apruzzi:2021nmk}
should be integrated (after refining it to differential cohomology) in
order to obtain the anomaly~\eqref{eq:1form-instAnom}: the effective
coupling induced by the second term in~\eqref{eq:CS_bulk_expanded} for
gauge fields $c=d\lambda$ is of the form
$d\lambda \wedge H \wedge d\lambda = d(\lambda \wedge H\wedge
d\lambda)$. If we identify the $\lambda$ with the RR fields $C$, this
is precisely the desired Wess-Zumino coupling on the boundary.

Nevertheless, it is also possible, and interesting, to verify
that~\eqref{eq:symm-inflow} does encode the
anomaly~\eqref{eq:1form-instAnom}. In order to do this it is not
sufficient to work at the level of differential forms, so we
reformulate the problem in terms of differential
cohomology. Accordingly, we uplift the field $c$ to the differential
cohomology field
$\diff a =\diff a_1+\diff a_3+ \diff a_5+\diff a_7+ \diff a_{9}+\diff
a_{11}$ with $\diff a_i\in \diff Z^{i}(\cN)$. Then, exactly as in the
case of the Chern-Simons theory we saw before, the first term uplifts
to
\[
 \frac{1}{2}\int_\cN \sh\left( \diff a \cdot \diff a^*\right)\, .
\]
The second term is harder to make globally well defined. A naive
uplift would send of the fields $c$ to $\diff a$, which results in a
differential cocycle of degree 12. This suggests that a proper
definition of the coupling requires us to further extend $\cN$ to a
12-dimensional manifold $\cW$ with boundary $\partial\cW=\cN$. Then,
the second term uplifts to
\[
  \label{eq:extended-coupling-action}
  -\int_\cW \sh\left( \diff a\cdot \diff H\cdot \diff a^*\right)\, .
\]
Note that, if we compare with~\eqref{eq:CS_bulk_expanded}, there is no
explicit factor of $1/2$ in this expression. To see that this is the
right normalisation, take $\diff a$ be top trivial, so we have
\[
\int_\cW h(\diff a_5\cdot \diff H\cdot \diff a_5)
&=\int_\cW h(\diff a_5)\wedge  R(\diff H)\wedge R(\diff a_5)\\
&=\int_\cW c_4 \wedge  R(\diff H)\wedge dc_4
= -\frac{1}{2} \int_\cW d(c_4 \wedge  R(\diff H)\wedge c_4)\\
&=
-\frac{1}{2} \int_\cN c_4 \wedge  R(\diff H)\wedge c_4\, .
\]

In our case, a simple choice for $\cW$, given that
$\cN = S^3/\bZ_n\times Y^8$, is to take
$\cW = \overline{\bC^2/\bZ_n}\times Y^8$, with
$\overline{\bC^2/\bZ_n}$ a compactification of $\bC^2/\bZ_n$, given
(for example) by the points of distance at most $1$ from the origin of
$\bC^2$ modulo $\bZ_n$. Here we are assuming that $Y^8$ is closed. We
should note that, in general, defining the extension $\cW$ goes
against our philosophy of working locally, and makes the analysis not
immediately applicable to the (most interesting case) where $Y^8$
itself has a boundary. As we will see, our analysis is local in $Y^8$,
so we will postulate that the right definition of the coupling is by
taking~\eqref{eq:extended-coupling-action}, with
$\cW = \overline{\bC^2/\bZ_n}\times Y^8$ also in the case that $Y^8$
has a boundary.

Under this assumption, using (\ref{eq:Stokes}), we write the refined
Chern-Simons action as
\[
    \mathsf{CS}_{H}[\diff a] = \sh\left( 
     \frac{1}{2}\, d \int_\cW \diff a \cdot \diff a^*- \int_\cW
     \diff a\cdot \diff H\cdot \diff a^*\right)\, .
\]
The anomaly term (\ref{eq:1form-instAnom}) results from the term
\[
\label{eq:a5ha5}
\mathsf{CS}_{H}[\diff a_5] = \int_\cW \sh\left( \diff H\cdot\diff a_5\cdot  \diff a_5\right)\, .
\]
More specifically in the symmetry inflow picture, we have
\[
  \delta\bigl( \cL_{\text{anomaly}}  \bigr) =
  2\pi i\, \sh\left(\int_{\overline{\bC^2/\bZ_n}} \Delta(\diff H\cdot\diff a_5\cdot \diff a_5))\right) \, .
\]
To show this, we perform the gauge transformations 
$\diff a_5 \to \diff a_5+d\diff b_{4}$ and find
\[
  \Delta(\diff H\cdot\diff a_5\cdot \diff a_5) = \diff H \cdot d\diff
  \lambda = - d(\diff H \cdot \diff \lambda) \, ,
\]
similarly to (\ref{eq:BFvariantion}), with
$\diff \lambda\df - \diff a_5\cdot \diff b_4+\diff b_4\cdot \diff
a_5-d\diff b_4\cdot \diff b_4$. We therefore have (by Stokes)
\[
  \int_{\overline{\bC^2/\bZ_n}} \Delta(\diff H\cdot\diff a_5\cdot
  \diff a_5))  =& -d \int_{\overline{\bC^2/\bZ_n}} \diff H \cdot \diff
  \lambda + \int_{S^3/\bZ_n} \diff H \cdot \diff \lambda \\
   =& -d(I(\diff H\cdot \diff \lambda)/{\overline{\bC^2/\bZ_n}}, \sh(\diff H\cdot \diff \lambda)/{\overline{\bC^2/\bZ_n}}, 0) + \int_{S^3/\bZ_n} \diff H\cdot \diff \lambda \\
   =& \,(-\delta I(\diff H\cdot \diff \lambda)/{\overline{\bC^2/\bZ_n}}, I(\diff H\cdot \diff \lambda)/{\overline{\bC^2/\bZ_n}} + \delta(\sh(\diff H\cdot \diff \lambda)/{\overline{\bC^2/\bZ_n}}), 0) \\ & +\int_{S^3/\bZ_n} \diff H \cdot \diff \lambda\, .
\]
From here
\[
  \label{eq:holonomy-filled-fiber}
  \sh\left(\int_{\overline{\bC^2/\bZ_n}} \Delta(\diff H\cdot\diff
    a_5\cdot \diff a_5))\right) = \delta
  \int_{\overline{\bC^2/\bZ_n}}\sh(\diff H\cdot \diff \lambda) +
  \int_{S^3/\bZ_n} \sh(\diff H\cdot \diff\lambda)\mod 1\, .
\]
To make further progress, we expand
$\diff b_4=\diff \beta_2\cdot \diff t$, with
$\diff \beta_2 = (\tilde c_2,r_2,0)$ and $\diff t=(t,\varphi,0)$ as in
section~\ref{sec:(1,1)-example}, and $\diff H=p^*\diff G$, with
$\diff G=(g, \gamma, \Gamma)$ a differential 3-cocycle on $X^7$
extending to $Y^8$,\footnote{Recall that our formalism, $H$-twisted
  K-theory, treats $H$ differently to the RR fields, and in particular
  it extends it \emph{classically} (namely, as a background 3-cocycle)
  into the bulk.} and $p\colon \cW\to Y^8$ the projection. We are
also, as usual, interested in pure gauge $\diff a_5$, so we will
choose $\diff a_5=\diff 0$ (any other pure gauge $\diff a_5$ is related to
this by a redefinition of $\diff b_4$). From
\[
  \diff H \cdot \diff \lambda = -(g\cup \tilde c_2\cup \delta \tilde
  c_2\cup t^2, g\cup \delta \tilde c_2\cup t \cup \tilde c_2\cup
  \varphi, 0)
\]
(where we have not indicated pullbacks explicitly) we then have that
the first term in~\eqref{eq:holonomy-filled-fiber} vanishes due to
degree reasons. Defining
\[
\ell \df\int_{S^3/\bZ_n} t\cup \varphi\, ,
\]
the second term gives:
\[
    \sh\left(\int_{\overline{\bC^2/\bZ_n}} \Delta(\diff H\cdot\diff
    a_5\cdot \diff a_5))\right) =-
    \ell\, g\cup \delta
  \tilde c_2\cup \tilde c_2  =-\frac{\ell}{2}\, \delta(g\cup \tilde c_2\cup \tilde c_2) \mod 1\, ,
\]
which is indeed of the expected form, with
\[
    n_{\text{inst}}(\tilde{c}_2)=-\frac{\ell}{2} \, \tilde{c}_2\cup \tilde{c}_2\mod 1\, .
\]

\acknowledgments

We thank Enrico Andriolo, Gabriel Arenas-Henriquez, Ibrahima Bah,
Federico Bonetti, Michele Del Zotto, Nabil Iqbal, Javier Magán, Sakura
Schäfer-Nameki, Tin Sulejmanpasic, Yuji Tachikawa, David Tong and
Kazuya Yonekura for helpful discussions. We are also very thankful to
the JHEP referee for a number of very useful comments and suggestions.
I.G.E. is partially supported by STFC grant ST/T000708/1 and by the
Simons Foundation collaboration grant 888990 on Global Categorical
Symmetries. S.S.H. is supported by WPI Initiative, MEXT, Japan at
Kavli IPMU, the University of Tokyo.

\paragraph{Data access statement.} There is no additional research
data associated with this work.

\bibliographystyle{JHEP}
\bibliography{refs}

\providecommand{\href}[2]{#2}\begingroup\raggedright\begin{thebibliography}{100}

\bibitem{Gaiotto:2010be}
D.~Gaiotto, G.~W. Moore, and A.~Neitzke, {\it {Framed BPS States}},  {\em Adv. Theor. Math. Phys.} {\bf 17} (2013), no.~2 241--397, [\href{http://arxiv.org/abs/1006.0146}{{\tt arXiv:1006.0146}}].

\bibitem{Aharony:2013hda}
O.~Aharony, N.~Seiberg, and Y.~Tachikawa, {\it {Reading between the lines of four-dimensional gauge theories}},  {\em JHEP} {\bf 08} (2013) 115, [\href{http://arxiv.org/abs/1305.0318}{{\tt arXiv:1305.0318}}].

\bibitem{Kapustin:2014gua}
A.~Kapustin and N.~Seiberg, {\it {Coupling a QFT to a TQFT and Duality}},  {\em JHEP} {\bf 04} (2014) 001, [\href{http://arxiv.org/abs/1401.0740}{{\tt arXiv:1401.0740}}].

\bibitem{Gaiotto:2014kfa}
D.~Gaiotto, A.~Kapustin, N.~Seiberg, and B.~Willett, {\it {Generalized Global Symmetries}},  {\em JHEP} {\bf 02} (2015) 172, [\href{http://arxiv.org/abs/1412.5148}{{\tt arXiv:1412.5148}}].

\bibitem{Frohlich:2009gb}
J.~Frohlich, J.~Fuchs, I.~Runkel, and C.~Schweigert, {\it {Defect lines, dualities, and generalised orbifolds}},  in {\em {16th International Congress on Mathematical Physics}}, 9, 2009.
\newblock \href{http://arxiv.org/abs/0909.5013}{{\tt arXiv:0909.5013}}.

\bibitem{Carqueville:2012dk}
N.~Carqueville and I.~Runkel, {\it {Orbifold completion of defect bicategories}},  {\em Quantum Topol.} {\bf 7} (2016), no.~2 203--279, [\href{http://arxiv.org/abs/1210.6363}{{\tt arXiv:1210.6363}}].

\bibitem{Brunner:2013xna}
I.~Brunner, N.~Carqueville, and D.~Plencner, {\it {A quick guide to defect orbifolds}},  {\em Proc. Symp. Pure Math.} {\bf 88} (2014) 231--242, [\href{http://arxiv.org/abs/1310.0062}{{\tt arXiv:1310.0062}}].

\bibitem{Bhardwaj:2017xup}
L.~Bhardwaj and Y.~Tachikawa, {\it {On finite symmetries and their gauging in two dimensions}},  {\em JHEP} {\bf 03} (2018) 189, [\href{http://arxiv.org/abs/1704.02330}{{\tt arXiv:1704.02330}}].

\bibitem{Gaiotto:2019xmp}
D.~Gaiotto and T.~Johnson-Freyd, {\it {Condensations in higher categories}},  \href{http://arxiv.org/abs/1905.09566}{{\tt arXiv:1905.09566}}.

\bibitem{Heidenreich:2021xpr}
B.~Heidenreich, J.~McNamara, M.~Montero, M.~Reece, T.~Rudelius, and I.~Valenzuela, {\it {Non-invertible global symmetries and completeness of the spectrum}},  {\em JHEP} {\bf 09} (2021) 203, [\href{http://arxiv.org/abs/2104.07036}{{\tt arXiv:2104.07036}}].

\bibitem{Choi:2021kmx}
Y.~Choi, C.~Cordova, P.-S. Hsin, H.~T. Lam, and S.-H. Shao, {\it {Noninvertible duality defects in 3+1 dimensions}},  {\em Phys. Rev. D} {\bf 105} (2022), no.~12 125016, [\href{http://arxiv.org/abs/2111.01139}{{\tt arXiv:2111.01139}}].

\bibitem{Kaidi:2021xfk}
J.~Kaidi, K.~Ohmori, and Y.~Zheng, {\it {Kramers-Wannier-like Duality Defects in (3+1)D Gauge Theories}},  {\em Phys. Rev. Lett.} {\bf 128} (2022), no.~11 111601, [\href{http://arxiv.org/abs/2111.01141}{{\tt arXiv:2111.01141}}].

\bibitem{Cordova:2022ruw}
C.~Cordova, T.~T. Dumitrescu, K.~Intriligator, and S.-H. Shao, {\it {Snowmass White Paper: Generalized Symmetries in Quantum Field Theory and Beyond}},  in {\em {Snowmass 2021}}, 5, 2022.
\newblock \href{http://arxiv.org/abs/2205.09545}{{\tt arXiv:2205.09545}}.

\bibitem{McGreevy:2022oyu}
J.~McGreevy, {\it {Generalized Symmetries in Condensed Matter}},  {\em Ann. Rev. Condensed Matter Phys.} {\bf 14} (2023) 57--82, [\href{http://arxiv.org/abs/2204.03045}{{\tt arXiv:2204.03045}}].

\bibitem{Freed:2022iao}
D.~S. Freed, {\it {Introduction to topological symmetry in QFT}},  \href{http://arxiv.org/abs/2212.00195}{{\tt arXiv:2212.00195}}.

\bibitem{Schafer-Nameki:2023jdn}
S.~Schafer-Nameki, {\it {ICTP lectures on (non-)invertible generalized symmetries}},  {\em Phys. Rept.} {\bf 1063} (2024) 1--55, [\href{http://arxiv.org/abs/2305.18296}{{\tt arXiv:2305.18296}}].

\bibitem{Brennan:2023mmt}
T.~D. Brennan and S.~Hong, {\it {Introduction to Generalized Global Symmetries in QFT and Particle Physics}},  \href{http://arxiv.org/abs/2306.00912}{{\tt arXiv:2306.00912}}.

\bibitem{Bhardwaj:2023kri}
L.~Bhardwaj, L.~E. Bottini, L.~Fraser-Taliente, L.~Gladden, D.~S.~W. Gould, A.~Platschorre, and H.~Tillim, {\it {Lectures on generalized symmetries}},  {\em Phys. Rept.} {\bf 1051} (2024) 1--87, [\href{http://arxiv.org/abs/2307.07547}{{\tt arXiv:2307.07547}}].

\bibitem{Shao:2023gho}
S.-H. Shao, {\it {What's Done Cannot Be Undone: TASI Lectures on Non-Invertible Symmetries}},  \href{http://arxiv.org/abs/2308.00747}{{\tt arXiv:2308.00747}}.

\bibitem{Ji:2019jhk}
W.~Ji and X.-G. Wen, {\it {Categorical symmetry and noninvertible anomaly in symmetry-breaking and topological phase transitions}},  {\em Phys. Rev. Res.} {\bf 2} (2020), no.~3 033417, [\href{http://arxiv.org/abs/1912.13492}{{\tt arXiv:1912.13492}}].

\bibitem{Gaiotto:2020iye}
D.~Gaiotto and J.~Kulp, {\it {Orbifold groupoids}},  {\em JHEP} {\bf 02} (2021) 132, [\href{http://arxiv.org/abs/2008.05960}{{\tt arXiv:2008.05960}}].

\bibitem{Apruzzi:2021nmk}
F.~Apruzzi, F.~Bonetti, I.~García~Etxebarria, S.~S. Hosseini, and S.~Schafer-Nameki, {\it {Symmetry TFTs from String Theory}},  {\em Commun. Math. Phys.} {\bf 402} (2023), no.~1 895--949, [\href{http://arxiv.org/abs/2112.02092}{{\tt arXiv:2112.02092}}].

\bibitem{Freed:2022qnc}
D.~S. Freed, G.~W. Moore, and C.~Teleman, {\it {Topological symmetry in quantum field theory}},  \href{http://arxiv.org/abs/2209.07471}{{\tt arXiv:2209.07471}}.

\bibitem{Kaidi:2022cpf}
J.~Kaidi, K.~Ohmori, and Y.~Zheng, {\it {Symmetry TFTs for Non-invertible Defects}},  {\em Commun. Math. Phys.} {\bf 404} (2023), no.~2 1021--1124, [\href{http://arxiv.org/abs/2209.11062}{{\tt arXiv:2209.11062}}].

\bibitem{Kaidi:2023maf}
J.~Kaidi, E.~Nardoni, G.~Zafrir, and Y.~Zheng, {\it {Symmetry TFTs and anomalies of non-invertible symmetries}},  {\em JHEP} {\bf 10} (2023) 053, [\href{http://arxiv.org/abs/2301.07112}{{\tt arXiv:2301.07112}}].

\bibitem{Bhardwaj:2023wzd}
L.~Bhardwaj and S.~Schafer-Nameki, {\it {Generalized Charges, Part I: Invertible Symmetries and Higher Representations}},  {\em SciPost Phys.} {\bf 16} (2024) 093, [\href{http://arxiv.org/abs/2304.02660}{{\tt arXiv:2304.02660}}].

\bibitem{Baume:2023kkf}
F.~Baume, J.~J. Heckman, M.~H\"ubner, E.~Torres, A.~P. Turner, and X.~Yu, {\it {SymTrees and Multi-Sector QFTs}},  \href{http://arxiv.org/abs/2310.12980}{{\tt arXiv:2310.12980}}.

\bibitem{Brennan:2024fgj}
T.~D. Brennan and Z.~Sun, {\it {A SymTFT for Continuous Symmetries}},  \href{http://arxiv.org/abs/2401.06128}{{\tt arXiv:2401.06128}}.

\bibitem{Antinucci:2024zjp}
A.~Antinucci and F.~Benini, {\it {Anomalies and gauging of U(1) symmetries}},  \href{http://arxiv.org/abs/2401.10165}{{\tt arXiv:2401.10165}}.

\bibitem{Bonetti:2024cjk}
F.~Bonetti, M.~Del~Zotto, and R.~Minasian, {\it {SymTFTs for Continuous non-Abelian Symmetries}},  \href{http://arxiv.org/abs/2402.12347}{{\tt arXiv:2402.12347}}.

\bibitem{Apruzzi:2024htg}
F.~Apruzzi, F.~Bedogna, and N.~Dondi, {\it {SymTh for non-finite symmetries}},  \href{http://arxiv.org/abs/2402.14813}{{\tt arXiv:2402.14813}}.

\bibitem{DelZotto:2024tae}
M.~Del~Zotto, S.~N. Meynet, and R.~Moscrop, {\it {Remarks on Geometric Engineering, Symmetry TFTs and Anomalies}},  \href{http://arxiv.org/abs/2402.18646}{{\tt arXiv:2402.18646}}.

\bibitem{Bhardwaj:2024qrf}
L.~Bhardwaj, D.~Pajer, S.~Schafer-Nameki, and A.~Warman, {\it {Hasse Diagrams for Gapless SPT and SSB Phases with Non-Invertible Symmetries}},  \href{http://arxiv.org/abs/2403.00905}{{\tt arXiv:2403.00905}}.

\bibitem{Cordova:2024vsq}
C.~Cordova, D.~Garc\'\i{}a-Sep\'ulveda, and N.~Holfester, {\it {Particle-Soliton Degeneracies from Spontaneously Broken Non-Invertible Symmetry}},  \href{http://arxiv.org/abs/2403.08883}{{\tt arXiv:2403.08883}}.

\bibitem{Nardoni:2024sos}
E.~Nardoni, M.~Sacchi, O.~Sela, G.~Zafrir, and Y.~Zheng, {\it {Dimensionally Reducing Generalized Symmetries from (3+1)-Dimensions}},  \href{http://arxiv.org/abs/2403.15995}{{\tt arXiv:2403.15995}}.

\bibitem{Argurio:2024oym}
R.~Argurio, F.~Benini, M.~Bertolini, G.~Galati, and P.~Niro, {\it {On the Symmetry TFT of Yang-Mills-Chern-Simons theory}},  \href{http://arxiv.org/abs/2404.06601}{{\tt arXiv:2404.06601}}.

\bibitem{Freed:2012bs}
D.~S. Freed and C.~Teleman, {\it {Relative quantum field theory}},  {\em Commun. Math. Phys.} {\bf 326} (2014) 459--476, [\href{http://arxiv.org/abs/1212.1692}{{\tt arXiv:1212.1692}}].

\bibitem{Bah:2019rgq}
I.~Bah, F.~Bonetti, R.~Minasian, and E.~Nardoni, {\it {Anomalies of QFTs from M-theory and Holography}},  {\em JHEP} {\bf 01} (2020) 125, [\href{http://arxiv.org/abs/1910.04166}{{\tt arXiv:1910.04166}}].

\bibitem{Bah:2020jas}
I.~Bah, F.~Bonetti, R.~Minasian, and P.~Weck, {\it {Anomaly Inflow Methods for SCFT Constructions in Type IIB}},  {\em JHEP} {\bf 02} (2021) 116, [\href{http://arxiv.org/abs/2002.10466}{{\tt arXiv:2002.10466}}].

\bibitem{GarciaEtxebarria:2019caf}
I.~García~Etxebarria, B.~Heidenreich, and D.~Regalado, {\it {IIB flux non-commutativity and the global structure of field theories}},  {\em JHEP} {\bf 10} (2019) 169, [\href{http://arxiv.org/abs/1908.08027}{{\tt arXiv:1908.08027}}].

\bibitem{Morrison:2020ool}
D.~R. Morrison, S.~Schafer-Nameki, and B.~Willett, {\it {Higher-Form Symmetries in 5d}},  {\em JHEP} {\bf 09} (2020) 024, [\href{http://arxiv.org/abs/2005.12296}{{\tt arXiv:2005.12296}}].

\bibitem{Albertini:2020mdx}
F.~Albertini, M.~Del~Zotto, I.~García~Etxebarria, and S.~S. Hosseini, {\it {Higher Form Symmetries and M-theory}},  {\em JHEP} {\bf 12} (2020) 203, [\href{http://arxiv.org/abs/2005.12831}{{\tt arXiv:2005.12831}}].

\bibitem{Moore:2004jv}
G.~W. Moore, {\it {Anomalies, Gauss laws, and Page charges in M-theory}},  {\em Comptes Rendus Physique} {\bf 6} (2005) 251--259, [\href{http://arxiv.org/abs/hep-th/0409158}{{\tt hep-th/0409158}}].

\bibitem{Freed:2006ya}
D.~S. Freed, G.~W. Moore, and G.~Segal, {\it {The Uncertainty of Fluxes}},  {\em Commun. Math. Phys.} {\bf 271} (2007) 247--274, [\href{http://arxiv.org/abs/hep-th/0605198}{{\tt hep-th/0605198}}].

\bibitem{Freed:2006yc}
D.~S. Freed, G.~W. Moore, and G.~Segal, {\it {Heisenberg Groups and Noncommutative Fluxes}},  {\em Annals Phys.} {\bf 322} (2007) 236--285, [\href{http://arxiv.org/abs/hep-th/0605200}{{\tt hep-th/0605200}}].

\bibitem{Witten:1996hc}
E.~Witten, {\it {Five-brane effective action in M theory}},  {\em J. Geom. Phys.} {\bf 22} (1997) 103--133, [\href{http://arxiv.org/abs/hep-th/9610234}{{\tt hep-th/9610234}}].

\bibitem{Witten:1998wy}
E.~Witten, {\it {AdS / CFT correspondence and topological field theory}},  {\em JHEP} {\bf 12} (1998) 012, [\href{http://arxiv.org/abs/hep-th/9812012}{{\tt hep-th/9812012}}].

\bibitem{Witten:1999vg}
E.~Witten, {\it {Duality relations among topological effects in string theory}},  {\em JHEP} {\bf 05} (2000) 031, [\href{http://arxiv.org/abs/hep-th/9912086}{{\tt hep-th/9912086}}].

\bibitem{Moore:1999gb}
G.~W. Moore and E.~Witten, {\it {Selfduality, Ramond-Ramond fields, and K theory}},  {\em JHEP} {\bf 05} (2000) 032, [\href{http://arxiv.org/abs/hep-th/9912279}{{\tt hep-th/9912279}}].

\bibitem{Freed:2000ta}
D.~S. Freed, {\it {Dirac charge quantization and generalized differential cohomology}},  \href{http://arxiv.org/abs/hep-th/0011220}{{\tt hep-th/0011220}}.

\bibitem{Gukov:2004id}
S.~Gukov, E.~Martinec, G.~W. Moore, and A.~Strominger, {\it {Chern-Simons gauge theory and the AdS(3) / CFT(2) correspondence}},  in {\em {From Fields to Strings: Circumnavigating Theoretical Physics: A Conference in Tribute to Ian Kogan}}, pp.~1606--1647, 3, 2004.
\newblock \href{http://arxiv.org/abs/hep-th/0403225}{{\tt hep-th/0403225}}.

\bibitem{Belov:2004ht}
D.~Belov and G.~W. Moore, {\it {Conformal blocks for AdS(5) singletons}},  \href{http://arxiv.org/abs/hep-th/0412167}{{\tt hep-th/0412167}}.

\bibitem{Belov:2006jd}
D.~Belov and G.~W. Moore, {\it {Holographic Action for the Self-Dual Field}},  \href{http://arxiv.org/abs/hep-th/0605038}{{\tt hep-th/0605038}}.

\bibitem{Belov:2006xj}
D.~M. Belov and G.~W. Moore, {\it {Type II Actions from 11-Dimensional Chern-Simons Theories}},  \href{http://arxiv.org/abs/hep-th/0611020}{{\tt hep-th/0611020}}.

\bibitem{Hsieh:2020jpj}
C.-T. Hsieh, Y.~Tachikawa, and K.~Yonekura, {\it {Anomaly Inflow and p-Form Gauge Theories}},  {\em Commun. Math. Phys.} {\bf 391} (2022), no.~2 495--608, [\href{http://arxiv.org/abs/2003.11550}{{\tt arXiv:2003.11550}}].

\bibitem{Hopkins:2002rd}
M.~J. Hopkins and I.~M. Singer, {\it {Quadratic functions in geometry, topology, and M theory}},  {\em J. Diff. Geom.} {\bf 70} (2005), no.~3 329--452, [\href{http://arxiv.org/abs/math/0211216}{{\tt math/0211216}}].

\bibitem{Camara:2011jg}
P.~G. Camara, L.~E. Ibanez, and F.~Marchesano, {\it {RR photons}},  {\em JHEP} {\bf 09} (2011) 110, [\href{http://arxiv.org/abs/1106.0060}{{\tt arXiv:1106.0060}}].

\bibitem{Casas:2023wlo}
G.~F. Casas, F.~Marchesano, and M.~Zatti, {\it {Torsion in cohomology and dimensional reduction}},  \href{http://arxiv.org/abs/2306.14959}{{\tt arXiv:2306.14959}}.

\bibitem{vanBeest:2022fss}
M.~van Beest, D.~S.~W. Gould, S.~Schafer-Nameki, and Y.-N. Wang, {\it {Symmetry TFTs for 3d QFTs from M-theory}},  {\em JHEP} {\bf 02} (2023) 226, [\href{http://arxiv.org/abs/2210.03703}{{\tt arXiv:2210.03703}}].

\bibitem{Apruzzi:2022rei}
F.~Apruzzi, I.~Bah, F.~Bonetti, and S.~Schafer-Nameki, {\it {Noninvertible Symmetries from Holography and Branes}},  {\em Phys. Rev. Lett.} {\bf 130} (2023), no.~12 121601, [\href{http://arxiv.org/abs/2208.07373}{{\tt arXiv:2208.07373}}].

\bibitem{Lawrie:2023tdz}
C.~Lawrie, X.~Yu, and H.~Y. Zhang, {\it {Intermediate Defect Groups, Polarization Pairs, and Non-invertible Duality Defects}},  \href{http://arxiv.org/abs/2306.11783}{{\tt arXiv:2306.11783}}.

\bibitem{Bah:2023ymy}
I.~Bah, E.~Leung, and T.~Waddleton, {\it {Non-invertible symmetries, brane dynamics, and tachyon condensation}},  {\em JHEP} {\bf 01} (2024) 117, [\href{http://arxiv.org/abs/2306.15783}{{\tt arXiv:2306.15783}}].

\bibitem{Apruzzi:2023uma}
F.~Apruzzi, F.~Bonetti, D.~S.~W. Gould, and S.~Schafer-Nameki, {\it {Aspects of Categorical Symmetries from Branes: SymTFTs and Generalized Charges}},  \href{http://arxiv.org/abs/2306.16405}{{\tt arXiv:2306.16405}}.

\bibitem{Yu:2023nyn}
X.~Yu, {\it {Non-invertible Symmetries in 2D from Type IIB String Theory}},  \href{http://arxiv.org/abs/2310.15339}{{\tt arXiv:2310.15339}}.

\bibitem{Basile:2023zng}
I.~Basile and G.~Leone, {\it {Anomaly constraints for heterotic strings and supergravity in six dimensions}},  {\em JHEP} {\bf 04} (2024) 067, [\href{http://arxiv.org/abs/2310.20480}{{\tt arXiv:2310.20480}}].

\bibitem{10.1007/BFb0075216}
J.~Cheeger and J.~Simons, {\it Differential characters and geometric invariants},  in {\em Geometry and Topology}, (Berlin, Heidelberg), pp.~50--80, Springer Berlin Heidelberg, 1985.

\bibitem{spanier1989algebraic}
E.~Spanier, {\em Algebraic Topology}.
\newblock McGraw-Hill series in higher mathematics. Springer, 1989.

\bibitem{Banks:2010zn}
T.~Banks and N.~Seiberg, {\it {Symmetries and Strings in Field Theory and Gravity}},  {\em Phys. Rev. D} {\bf 83} (2011) 084019, [\href{http://arxiv.org/abs/1011.5120}{{\tt arXiv:1011.5120}}].

\bibitem{Kravec:2013pua}
S.~M. Kravec and J.~McGreevy, {\it {A gauge theory generalization of the fermion-doubling theorem}},  {\em Phys. Rev. Lett.} {\bf 111} (2013) 161603, [\href{http://arxiv.org/abs/1306.3992}{{\tt arXiv:1306.3992}}].

\bibitem{Kravec:2014aza}
S.~M. Kravec, J.~McGreevy, and B.~Swingle, {\it {All-fermion electrodynamics and fermion number anomaly inflow}},  {\em Phys. Rev. D} {\bf 92} (2015), no.~8 085024, [\href{http://arxiv.org/abs/1409.8339}{{\tt arXiv:1409.8339}}].

\bibitem{Wang:2018qoy}
J.~Wang, X.-G. Wen, and E.~Witten, {\it {A New SU(2) Anomaly}},  {\em J. Math. Phys.} {\bf 60} (2019), no.~5 052301, [\href{http://arxiv.org/abs/1810.00844}{{\tt arXiv:1810.00844}}].

\bibitem{Seiberg:2018ntt}
N.~Seiberg, Y.~Tachikawa, and K.~Yonekura, {\it {Anomalies of Duality Groups and Extended Conformal Manifolds}},  {\em PTEP} {\bf 2018} (2018), no.~7 073B04, [\href{http://arxiv.org/abs/1803.07366}{{\tt arXiv:1803.07366}}].

\bibitem{Hsieh:2019iba}
C.-T. Hsieh, Y.~Tachikawa, and K.~Yonekura, {\it {Anomaly of the Electromagnetic Duality of Maxwell Theory}},  {\em Phys. Rev. Lett.} {\bf 123} (2019), no.~16 161601, [\href{http://arxiv.org/abs/1905.08943}{{\tt arXiv:1905.08943}}].

\bibitem{DelZotto:2022ras}
M.~Del~Zotto and I.~García~Etxebarria, {\it {Global Structures from the Infrared}},  \href{http://arxiv.org/abs/2204.06495}{{\tt arXiv:2204.06495}}.

\bibitem{Gukov:1998kn}
S.~Gukov, M.~Rangamani, and E.~Witten, {\it {Dibaryons, strings and branes in AdS orbifold models}},  {\em JHEP} {\bf 12} (1998) 025, [\href{http://arxiv.org/abs/hep-th/9811048}{{\tt hep-th/9811048}}].

\bibitem{Hatcher}
A.~Hatcher, {\em Algebraic Topology}.
\newblock Algebraic Topology. Cambridge University Press, 2002.

\bibitem{Buscher:1987qj}
T.~H. Buscher, {\it {Path Integral Derivation of Quantum Duality in Nonlinear Sigma Models}},  {\em Phys. Lett. B} {\bf 201} (1988) 466--472.

\bibitem{Rocek:1991ps}
M.~Rocek and E.~P. Verlinde, {\it {Duality, quotients, and currents}},  {\em Nucl. Phys. B} {\bf 373} (1992) 630--646, [\href{http://arxiv.org/abs/hep-th/9110053}{{\tt hep-th/9110053}}].

\bibitem{Schwarz:1993vs}
J.~H. Schwarz and A.~Sen, {\it {Duality symmetric actions}},  {\em Nucl. Phys. B} {\bf 411} (1994) 35--63, [\href{http://arxiv.org/abs/hep-th/9304154}{{\tt hep-th/9304154}}].

\bibitem{Seiberg:1994rs}
N.~Seiberg and E.~Witten, {\it {Electric - magnetic duality, monopole condensation, and confinement in N=2 supersymmetric Yang-Mills theory}},  {\em Nucl. Phys. B} {\bf 426} (1994) 19--52, [\href{http://arxiv.org/abs/hep-th/9407087}{{\tt hep-th/9407087}}]. [Erratum: Nucl.Phys.B 430, 485--486 (1994)].

\bibitem{Witten:1995gf}
E.~Witten, {\it {On S duality in Abelian gauge theory}},  {\em Selecta Math.} {\bf 1} (1995) 383, [\href{http://arxiv.org/abs/hep-th/9505186}{{\tt hep-th/9505186}}].

\bibitem{Villain:1974ir}
J.~Villain, {\it {Theory of one-dimensional and two-dimensional magnets with an easy magnetization plane. 2. The Planar, classical, two-dimensional magnet}},  {\em J. Phys. (France)} {\bf 36} (1975) 581--590.

\bibitem{Anosova:2022cjm}
M.~Anosova, C.~Gattringer, and T.~Sulejmanpasic, {\it {Self-dual U(1) lattice field theory with a \ensuremath{\theta}-term}},  {\em JHEP} {\bf 04} (2022) 120, [\href{http://arxiv.org/abs/2201.09468}{{\tt arXiv:2201.09468}}].

\bibitem{Maldacena:2001ss}
J.~M. Maldacena, G.~W. Moore, and N.~Seiberg, {\it {D-brane charges in five-brane backgrounds}},  {\em JHEP} {\bf 10} (2001) 005, [\href{http://arxiv.org/abs/hep-th/0108152}{{\tt hep-th/0108152}}].

\bibitem{Tong:2022gpg}
D.~Tong, {\it {A gauge theory for shallow water}},  {\em SciPost Phys.} {\bf 14} (2023), no.~5 102, [\href{http://arxiv.org/abs/2209.10574}{{\tt arXiv:2209.10574}}].

\bibitem{Losev:1995cr}
A.~Losev, G.~W. Moore, N.~Nekrasov, and S.~Shatashvili, {\it {Four-dimensional avatars of two-dimensional RCFT}},  {\em Nucl. Phys. B Proc. Suppl.} {\bf 46} (1996) 130--145, [\href{http://arxiv.org/abs/hep-th/9509151}{{\tt hep-th/9509151}}].

\bibitem{Moore:1989yh}
G.~W. Moore and N.~Seiberg, {\it {Taming the Conformal Zoo}},  {\em Phys. Lett. B} {\bf 220} (1989) 422--430.

\bibitem{Costello:2019tri}
K.~Costello and M.~Yamazaki, {\it {Gauge Theory And Integrability, III}},  \href{http://arxiv.org/abs/1908.02289}{{\tt arXiv:1908.02289}}.

\bibitem{Floreanini:1987as}
R.~Floreanini and R.~Jackiw, {\it {Selfdual Fields as Charge Density Solitons}},  {\em Phys. Rev. Lett.} {\bf 59} (1987) 1873.

\bibitem{Henneaux:1988gg}
M.~Henneaux and C.~Teitelboim, {\it {Dynamics of Chiral (Selfdual) $P$ Forms}},  {\em Phys. Lett. B} {\bf 206} (1988) 650--654.

\bibitem{Perry:1996mk}
M.~Perry and J.~H. Schwarz, {\it {Interacting chiral gauge fields in six-dimensions and Born-Infeld theory}},  {\em Nucl. Phys. B} {\bf 489} (1997) 47--64, [\href{http://arxiv.org/abs/hep-th/9611065}{{\tt hep-th/9611065}}].

\bibitem{Pasti:1996vs}
P.~Pasti, D.~P. Sorokin, and M.~Tonin, {\it {On Lorentz invariant actions for chiral p forms}},  {\em Phys. Rev. D} {\bf 55} (1997) 6292--6298, [\href{http://arxiv.org/abs/hep-th/9611100}{{\tt hep-th/9611100}}].

\bibitem{Pasti:1997gx}
P.~Pasti, D.~P. Sorokin, and M.~Tonin, {\it {Covariant action for a D = 11 five-brane with the chiral field}},  {\em Phys. Lett. B} {\bf 398} (1997) 41--46, [\href{http://arxiv.org/abs/hep-th/9701037}{{\tt hep-th/9701037}}].

\bibitem{Buratti:2019guq}
G.~Buratti, K.~Lechner, and L.~Melotti, {\it {Self-interacting chiral p-forms in higher dimensions}},  {\em Phys. Lett. B} {\bf 798} (2019) 135018, [\href{http://arxiv.org/abs/1909.10404}{{\tt arXiv:1909.10404}}].

\bibitem{Sen:2019qit}
A.~Sen, {\it {Self-dual forms: Action, Hamiltonian and Compactification}},  {\em J. Phys. A} {\bf 53} (2020), no.~8 084002, [\href{http://arxiv.org/abs/1903.12196}{{\tt arXiv:1903.12196}}].

\bibitem{Lambert:2023qgs}
N.~Lambert, {\it {Duality and fluxes in the sen formulation of self-dual fields}},  {\em Phys. Lett. B} {\bf 840} (2023) 137888, [\href{http://arxiv.org/abs/2302.10955}{{\tt arXiv:2302.10955}}].

\bibitem{Hull:2023dgp}
C.~M. Hull, {\it {Covariant action for self-dual p-form gauge fields in general spacetimes}},  {\em JHEP} {\bf 04} (2024) 011, [\href{http://arxiv.org/abs/2307.04748}{{\tt arXiv:2307.04748}}].

\bibitem{Avetisyan:2022zza}
Z.~Avetisyan, O.~Evnin, and K.~Mkrtchyan, {\it {Nonlinear (chiral) p-form electrodynamics}},  {\em JHEP} {\bf 08} (2022) 112, [\href{http://arxiv.org/abs/2205.02522}{{\tt arXiv:2205.02522}}].

\bibitem{Evnin:2022kqn}
O.~Evnin and K.~Mkrtchyan, {\it {Three approaches to chiral form interactions}},  {\em Differ. Geom. Appl.} {\bf 89} (2023) 102016, [\href{http://arxiv.org/abs/2207.01767}{{\tt arXiv:2207.01767}}].

\bibitem{Evnin:2023ypu}
O.~Evnin, E.~Joung, and K.~Mkrtchyan, {\it {Democratic Lagrangians from topological bulk}},  {\em Phys. Rev. D} {\bf 109} (2024), no.~6 066003, [\href{http://arxiv.org/abs/2309.04625}{{\tt arXiv:2309.04625}}].

\bibitem{Whitney}
H.~Whitney, {\em Geometric Integration Theory}.
\newblock Princeton University Press, 1957.

\bibitem{Wilson}
S.~O. Wilson, {\it Cochain algebra on manifolds and convergence under refinement},  {\em Topology and its Applications} {\bf 154} (2007), no.~9 1898--1920.

\bibitem{Pulmann:2019vrw}
J.~Pulmann, P.~\v{S}evera, and F.~Valach, {\it {A nonabelian duality for (higher) gauge theories}},  {\em Adv. Theor. Math. Phys.} {\bf 25} (2021), no.~1 241--274, [\href{http://arxiv.org/abs/1909.06151}{{\tt arXiv:1909.06151}}].

\bibitem{Avetisyan:2021heg}
Z.~Avetisyan, O.~Evnin, and K.~Mkrtchyan, {\it {Democratic Lagrangians for Nonlinear Electrodynamics}},  {\em Phys. Rev. Lett.} {\bf 127} (2021), no.~27 271601, [\href{http://arxiv.org/abs/2108.01103}{{\tt arXiv:2108.01103}}].

\bibitem{Fliss:2023dze}
J.~R. Fliss and S.~Vitouladitis, {\it {Entanglement in BF theory I: Essential topological entanglement}},  \href{http://arxiv.org/abs/2306.06158}{{\tt arXiv:2306.06158}}.

\bibitem{Fliss:2023uiv}
J.~R. Fliss and S.~Vitouladitis, {\it {Entanglement in BF theory II: Edge-modes}},  \href{http://arxiv.org/abs/2310.18391}{{\tt arXiv:2310.18391}}.

\bibitem{Douglas:2014ywa}
M.~R. Douglas, D.~S. Park, and C.~Schnell, {\it {The Cremmer-Scherk Mechanism in F-theory Compactifications on K3 Manifolds}},  {\em JHEP} {\bf 05} (2014) 135, [\href{http://arxiv.org/abs/1403.1595}{{\tt arXiv:1403.1595}}].

\bibitem{Roumpedakis:2022aik}
K.~Roumpedakis, S.~Seifnashri, and S.-H. Shao, {\it {Higher Gauging and Non-invertible Condensation Defects}},  {\em Commun. Math. Phys.} {\bf 401} (2023), no.~3 3043--3107, [\href{http://arxiv.org/abs/2204.02407}{{\tt arXiv:2204.02407}}].

\bibitem{Choi:2022zal}
Y.~Choi, C.~Cordova, P.-S. Hsin, H.~T. Lam, and S.-H. Shao, {\it {Non-invertible Condensation, Duality, and Triality Defects in 3+1 Dimensions}},  {\em Commun. Math. Phys.} {\bf 402} (2023), no.~1 489--542, [\href{http://arxiv.org/abs/2204.09025}{{\tt arXiv:2204.09025}}].

\bibitem{GarciaEtxebarria:2022vzq}
I.~García~Etxebarria, {\it {Branes and Non-Invertible Symmetries}},  {\em Fortsch. Phys.} {\bf 70} (2022), no.~11 2200154, [\href{http://arxiv.org/abs/2208.07508}{{\tt arXiv:2208.07508}}].

\bibitem{Heckman:2022muc}
J.~J. Heckman, M.~H\"ubner, E.~Torres, and H.~Y. Zhang, {\it {The Branes Behind Generalized Symmetry Operators}},  {\em Fortsch. Phys.} {\bf 71} (2023), no.~1 2200180, [\href{http://arxiv.org/abs/2209.03343}{{\tt arXiv:2209.03343}}].

\bibitem{Heckman:2022xgu}
J.~J. Heckman, M.~Hubner, E.~Torres, X.~Yu, and H.~Y. Zhang, {\it {Top down approach to topological duality defects}},  {\em Phys. Rev. D} {\bf 108} (2023), no.~4 046015, [\href{http://arxiv.org/abs/2212.09743}{{\tt arXiv:2212.09743}}].

\bibitem{Etheredge:2023ler}
M.~Etheredge, I.~García~Etxebarria, B.~Heidenreich, and S.~Rauch, {\it {Branes and symmetries for $ \mathcal{N} $ = 3 S-folds}},  {\em JHEP} {\bf 09} (2023) 005, [\href{http://arxiv.org/abs/2302.14068}{{\tt arXiv:2302.14068}}].

\bibitem{Dierigl:2023jdp}
M.~Dierigl, J.~J. Heckman, M.~Montero, and E.~Torres, {\it {R7-branes as charge conjugation operators}},  {\em Phys. Rev. D} {\bf 109} (2024), no.~4 046004, [\href{http://arxiv.org/abs/2305.05689}{{\tt arXiv:2305.05689}}].

\bibitem{Donagi:2023sbk}
R.~Donagi and M.~Wijnholt, {\it {The $M$-Theory Three-Form and Singular Geometries}},  \href{http://arxiv.org/abs/2310.05838}{{\tt arXiv:2310.05838}}.

\bibitem{Tachikawa:2013hya}
Y.~Tachikawa, {\it {On the 6d origin of discrete additional data of 4d gauge theories}},  {\em JHEP} {\bf 05} (2014) 020, [\href{http://arxiv.org/abs/1309.0697}{{\tt arXiv:1309.0697}}].

\bibitem{Monnier:2017klz}
S.~Monnier, {\it {The anomaly field theories of six-dimensional (2,0) superconformal theories}},  {\em Adv. Theor. Math. Phys.} {\bf 22} (2018) 2035--2089, [\href{http://arxiv.org/abs/1706.01903}{{\tt arXiv:1706.01903}}].

\bibitem{DelZotto:2015isa}
M.~Del~Zotto, J.~J. Heckman, D.~S. Park, and T.~Rudelius, {\it {On the Defect Group of a 6D SCFT}},  {\em Lett. Math. Phys.} {\bf 106} (2016), no.~6 765--786, [\href{http://arxiv.org/abs/1503.04806}{{\tt arXiv:1503.04806}}].

\bibitem{Fiorenza:2019usl}
D.~Fiorenza, H.~Sati, and U.~Schreiber, {\it {Twisted Cohomotopy implies M-theory anomaly cancellation on 8-manifolds}},  {\em Commun. Math. Phys.} {\bf 377} (2020), no.~3 1961--2025, [\href{http://arxiv.org/abs/1904.10207}{{\tt arXiv:1904.10207}}].

\bibitem{Evslin:2006cj}
J.~Evslin, {\it {What does(n't) K-theory classify?}},  \href{http://arxiv.org/abs/hep-th/0610328}{{\tt hep-th/0610328}}.

\bibitem{Witten:2000nv}
E.~Witten, {\it {Supersymmetric index in four-dimensional gauge theories}},  {\em Adv. Theor. Math. Phys.} {\bf 5} (2002) 841--907, [\href{http://arxiv.org/abs/hep-th/0006010}{{\tt hep-th/0006010}}].

\bibitem{Apruzzi:2020zot}
F.~Apruzzi, M.~Dierigl, and L.~Lin, {\it {The fate of discrete 1-form symmetries in 6d}},  {\em SciPost Phys.} {\bf 12} (2022), no.~2 047, [\href{http://arxiv.org/abs/2008.09117}{{\tt arXiv:2008.09117}}].

\bibitem{Freed:2009mw}
D.~S. Freed and J.~Lott, {\it {An index theorem in differential $K$\textendash{}theory}},  {\em Geom. Topol.} {\bf 14} (2010), no.~2 903--966, [\href{http://arxiv.org/abs/0907.3508}{{\tt arXiv:0907.3508}}].

\bibitem{lott_1994}
J.~Lott, {\it $\mathbf{R/Z}$ index theory},  {\em Communications in Analysis and Geometry} {\bf 2} (1994), no.~2 279--311.

\bibitem{simons2008structured}
J.~Simons and D.~Sullivan, {\it Structured vector bundles define differential k-theory},  2008.

\bibitem{tradler2012elementary}
T.~Tradler, S.~O. Wilson, and M.~Zeinalian, {\it An elementary differential extension of odd k-theory},  2012.

\bibitem{bunke2009smooth}
U.~Bunke and T.~Schick, {\it Smooth k-theory},  2009.

\bibitem{gorokhovsky2018hilbert}
A.~Gorokhovsky and J.~Lott, {\it A hilbert bundle description of differential k-theory},  2018.

\bibitem{Cvetic:2021sxm}
M.~Cvetic, M.~Dierigl, L.~Lin, and H.~Y. Zhang, {\it {Higher-form symmetries and their anomalies in M-/F-theory duality}},  {\em Phys. Rev. D} {\bf 104} (2021), no.~12 126019, [\href{http://arxiv.org/abs/2106.07654}{{\tt arXiv:2106.07654}}].

\end{thebibliography}\endgroup

\end{document}